\documentclass[%
 aip,
 amsmath,amssymb,
 reprint,%
jcp]{revtex4-1}

\usepackage{graphicx}% Include figure files
\usepackage{dcolumn}% Align table columns on decimal point
\usepackage{bm}% bold math
\graphicspath{{Figures/}} 
\usepackage[utf8]{inputenc}
\usepackage[T1]{fontenc}
\usepackage{mathptmx}
\usepackage{hyperref}
\usepackage{xcolor}

\begin{document}

\preprint{AIP/123-QED}

\title[Waiting-time dependent non-equilibrium phase diagram of simple glass- and gel-forming liquids]{Waiting-time dependent non-equilibrium phase diagram of  \\  simple glass- and gel-forming liquids}
% Force line breaks with \\

\author{Jes\'us Benigno Zepeda-L\'opez}
\email{benignoz@ifisica.uaslp.mx} 
 
\author{Magdaleno Medina-Noyola}%
 
\affiliation{ 
Instituto de F\'{\i}sica {\sl ``Manuel Sandoval Vallarta"},
Universidad Aut\'{o}noma de San Luis Potos\'{\i}, \'{A}lvaro
Obreg\'{o}n 64, 78000 San Luis Potos\'{\i}, SLP, M\'{e}xico.%\\This line break forced with \textbackslash\textbackslash
}%

\date{\today}% It is always \today, today,
             %  but any date may be explicitly specified

\begin{abstract}
Under many circumstances many soft and hard materials are present in a puzzling wealth of non-equilibrium amorphous states, whose properties are not stationary and depend on preparation. They are often summarized in unconventional ``phase diagrams'' that exhibit new ``phases'' and/or ``transitions'', in which the time, however, is an essential variable. This work proposes a solution to the problem of theoretically defining and predicting these non-equilibrium phases and their time-evolving phase diagrams, given the underlying molecular interactions. We  demonstrate that these non-equilibrium phases and the corresponding non-stationary (i.e., aging) phase diagrams can indeed be defined and predicted using the kinetic perspective of a novel non-equilibrium statistical mechanical theory of irreversible processes. This is illustrated with the theoretical description of the  transient process of dynamic arrest into non-equilibrium \emph{amorphous} solid phases, of  an instantaneously-quenched simple model fluid involving repulsive hard-sphere plus attractive square well pair interactions.
\end{abstract}

\maketitle

\section{\label{sect1} Introduction}

The formation of many soft and hard materials from the cooling of liquids and melts generates fascinating  non-equilibrium transient states. These  include glasses and gels, represented in rather unorthodox ``phase diagrams''  in which the (``waiting'') time after preparation becomes a key variable. Examples range from gel or glass formation by clays \cite{ruzicka1,ruzicka2} or proteins \cite{cardinaux,gibaud} in  solution, to the experimental phase transformations in the manufacture of hard materials, such as iron and steel \cite{callisterrethwisch,atlasttd} and porous glasses \cite{nakashima}. Ideally, these irreversible processes, and the non-equilibrium states involved, should be understood and \emph{predicted} as a direct consequence of general  physical principles, just like  thermodynamic equilibrium phases and phase transitions can be explained in terms of the underlying molecular interactions by the maximum-entropy principle \cite{callen}, together with Boltzmann's universal expression for the entropy, $S=k_B\ln W$, in terms of the number $W$ of microscopic states \cite{huang,mcquarrie,hansen}.  A simple and intuitive use of these principles is illustrated by van der Waals explanation of the origin of the gas and liquid equilibrium phases and their coexistence \cite{vdw,widom}.

Unfortunately, it is not common knowledge how Boltzmann's principle operates in general to govern the formation of these ubiquitous \emph{non-equilibrium} amorphous materials, whose properties do depend on time and preparation and whose fundamental understanding  is some times referred to as ``the deepest and most interesting unsolved problem in solid state theory'' \cite{anderson}. The main purpose of this work is to demonstrate that a sound theoretical definition, and first-principles prediction, of the concept of non-equilibrium phases and time-dependent (i.e., aging) phase diagrams, may be provided by  the kinetic perspective of a novel non-equilibrium statistical mechanical theory of irreversible processes in liquids, which results from the merging of Boltzmann's postulates with Onsager's fundamental principles of transport and fluctuations \cite{nescgle1}.

The rich and diverse phenomenology exhibited during the  irreversible formation of these  non-equilibrium amorphous materials is represented by an exploding body of experimental\cite{angellreview1,edigerreview1,ngaireview1,mckennasimon} and simulation \cite{sciortinotartaglia2005,royalturci,hunterweeks,Zaccarelli2005,Zaccarelli2007,Kob2010,Coniglio2014} data.  As a dramatic illustrative example, let us consider a simple liquid quenched from supercritical temperatures to inside the spinodal region, which triggers the irreversible process of spinodal decomposition. This process starts with the spatial amplification of unstable density fluctuations and is expected to end in the full  liquid-gas  phase separation of the originally uniform fluid \cite{cahnhilliard,cook,furukawa,langer,dhont,goryachev}. In certain colloidal liquids, however, this process  is interrupted \cite{cardinaux,gibaud,luetalnature,sanz,espinosafrisken,Gao,foffi2014} when the particles cluster in a percolating network, forming an amorphous sponge-like non-equilibrium bicontinuous structure,  typical of physical gels \cite{zaccarellireviewgels}. 

Similar spinodal decomposition phenomena occur in binary colloidal mixtures with short-range attraction, in which arrested microphase demixing leads to the formation of bi-gels \cite{foffi2014} and of complex gels formed by clusters of immobile particles surrounded by particles of the other species that remain mobile \cite{harden}. Spinodal demixing is also observed in many organic and inorganic materials including polymers, metallic alloys, and ceramics \cite{craievich1}, thus providing a route to the fabrication of, for example, silicate and borate porous glasses  \cite{craievich2}. In understanding these experimental observations, computer simulations in well-defined model systems have also
played an essential role. This includes Brownian dynamics (BD) \cite{heyeslodge} and molecular dynamics (MD) \cite{testardjcp}
investigations of the density- and temperature-dependence of the liquid-gas phase separation kinetics in suddenly-quenched
Lennard-Jones and square-well \cite{foffi2002,foffi2005} liquids. It also includes BD simulation of gel formation in the primitive model of electrolytes \cite{sanz}, and MD  simulations of bigel formation driven by demixing \cite{foffi2014}. 

The theoretical approach to the same fundamental challenge has also been the subject of a rich and well-documented discussion \cite{berthierreview,berthierbook2011}. Still, there are many loose ends in this phenomenological puzzle, whose full understanding calls for a canonical non-equilibrium statistical thermodynamic formalism that  incorporates time and irreversibility from the very beginning and explains this phenomenology from first-principles, i.e.,  in terms of interparticle interactions.  To a large extent this program started with the successful predictions of mode coupling theory (MCT)  of dynamic arrest  \cite{goetze1,goetze2,goetze3,goetze4,mctjenssen}, which stands out by its pioneering first-principles description of liquids near their glass transition, with many of its predictions finding a remarkable verification in  the observed features of the initial slowdown of real \cite{vanmegenpusey,vanmegenmortensen} and simulated \cite{kobandersen} supercooled liquids. Some of the general MCT ideas were also cast in the theory referred to as random first-order theory \cite{kinkpatrickwolynes} (RFOT), although with more limited interest on ``the hardest part of the detailed microscopic modeling'' \cite{lubchenkowolynes}. An illustrative example of the valuable specific predictive power of MCT  was its prediction of the existence of repulsive and attractive glasses in hard-colloids with short-range attractions \cite{fabian,bergenholtzfuchs,foffi2002}, including the early MCT predictions on the nature of the critical decay at higher-order glass-transition singularities in systems with short-ranged attractive potentials \cite{dawsonpre2001,goetzesperl1,goetzesperl2} , and the dynamics of colloidal liquids near a crossing of glass- and gel-transition lines \cite{sperl1}, whose validity was demonstrated experimentally \cite{eckert,phamscience2002,mallamace2000,wrchen} and by simulations \cite{foffidawson2002,foffidawson2002,sciortinotartaglia2005}.

In the first decade of this century an alternative theory, referred to as the  self-consistent generalized Langevin equation (SCGLE) theory of colloid dynamics \cite{scgle3,scgle4} and dynamic arrest \cite{todos1,pedroatractivos}, joined MCT in this effort, as reviewed in detail in Ref. \cite{reviewroque}. In spite of the noticeable differences in their fundamental origin and in the details of the approximations involved, this theory turned out to be in many respects remarkably similar to MCT \cite{goryvoigtmann}. MCT and the SCGLE theory are, however, unable to describe irreversible non-stationary processes such as the aging typical of glassy amorphous materials. Furthermore, they predict a divergence of the $\alpha$-relaxation time $\tau_\alpha$ at a critical temperature $T_c$, which is never observed in practice in molecular systems \cite{angellreview1,edigerreview1, ngaireview1}. The reason is that both theories are in reality intrinsically limited to describe the dynamics of liquids in their thermodynamic equilibrium states. 

The earliest attempts to surmount this fundamental limitation in the framework of MCT were made by Latz \cite{latz1,latz2} in 2000, who employed non-equilibrium projection operators to derive a set of exact equations that govern the coupled evolution of the non-stationary structure factor $S(k;t)$ and intermediate scattering functions $F(k,\tau; t)$ and $F_S(k,\tau; t)$, in terms of two-time memory functions, thus proposing a formal extension of MCT to situations far away from equilibrium. Latz's attention was also focused \cite{latz3} on the possible relation of his equations with the  mean-field theory of the aging dynamics of simple spin glass models \cite{crisanti1,crisanti2}, which had already made relevant and verifiable predictions  \cite{cugliandolo1,cugliandolo2}.  These models, however, lack a geometric structure and hence, cannot describe the spatial structure of real \cite{angellreview1} or simulated \cite{hunterweeks} \emph{structural} glass formers. Unfortunately, no concrete application of Latz's proposal to specific  examples of structural glasses is available in the literature. To complete this brief review, let us also mention the discussion on time-translational invariance and the fluctuation-dissipation theorem in the context of the description of slow dynamics in system out of equilibrium but close to dynamical arrest by De Gregorio et al. \cite{degregorio}. Similarly, we should also mention that Chen and Schweizer addressed aging based on a particle- and force-level theory that builds on MCT but including  activated dynamics to describe the coupled aging dynamics and the non-equilibrium evolution of the structure factor \cite{chenwchweitzer1,chenwchweitzer2,chenwchweitzer3,chenwchweitzer4}. 

In this context, in 2010 a generic theory of irreversible processes in liquids  was derived  \cite{nescgle0,nescgle1} from the assumption that the manner in which Boltzmann's postulate explains non-equilibrium states, is provided by a spatially non-local and temporally  non-Markovian and non-stationary generalization \cite{nescgle0} of Onsager's theory of thermal fluctuations \cite{onsagermachlup1, onsagermachlup2}. 
%(previously extended to allow for memory effects and spatial non-locality \cite{delrio, faraday}). 
For simple liquids with purely repulsive interactions this theory, referred to as the non-equilibrium self-consistent generalized Langevin equation (NE-SCGLE) theory \cite{nescgle1,nescgle3}, provided a detailed description of the  non-stationary and non-equilibrium transformation of equilibrium hard- (and soft)-sphere liquids, into ``repulsive'' (high-temperature, high-density) hard-sphere glasses \cite{nescgle3,nescgle4}. These predictions have recently found a reasonable agreement with the corresponding non-equilibrium simulations  \cite{nescgle6},  naturally explaining some of the most essential signatures of the glass transition \cite{angellreview1,edigerreview1,ngaireview1}. This success has rapidly branched in several relevant and surprising directions.

In the absence yet of a detailed review, let us briefly describe these recent developments. For example, when adding attractions to the purely repulsive interactions just referred to, the same theory predicts new dynamically-arrested phases, identified with gels and porous glasses \cite{nescgle5}, and provides a kinetic perspective of the irreversible evolution of the structure of the system after being instantaneously quenched to the interior of its spinodal region \cite{nescgle7,nescgle8}. It also explains the otherwise seemingly complex interplay between spinodal decomposition, gelation, glass transition, and their combinations \cite{sastry, cocard, wyss, zaconedelgado, guoleheny, coniglio, chaudhuri2}. Extended to  multi-component systems \cite{nescgle4,nescgle6}, for example, the NE-SCGLE theory opens the possibility of describing the aging of  ``double'' and ``single'' glasses in mixtures of neutral \cite{voigtmanndoubleglasses,rigo,lazaro} and charged \cite{prlrigoluis,portadajcppedro} particles; the initial steps in this direction are highly encouraging \cite{expsimbinarymixt2017}. Similarly, its extension to liquids formed by particles interacting by non-radially symmetric forces \cite{gory1,gory2}, accurately predicts the non-equilibrium coupled translational and rotational dynamic arrest observed in simulations \cite{gory3}.  

Under these circumstances, anyone interested in the statistical physics of liquids, may raise a natural and immediate question: what is the relationship, and possible interplay, between the conventional statistical thermodynamic theory of equilibrium fluids  (represented, for example, by integral equation \cite{mcquarrie,hansen} or density functional theory \cite{evans}) and this emerging non-equilibrium theory?.  We are convinced that such relationship will be intimate enough to stimulate  a creative communication between the practitioners of these two areas of statistical physics. One explicit example of common interest is, of course, the challenge of determining the Helmholtz free energy density functional $\mathcal{F}[n,T]$ in any new application. Determining $\mathcal{F}[n,T]$ is a classical and relevant  \emph{goal} of equilibrium statistical thermodynamics, while $\mathcal{F}[n,T]$ is an essential \emph{input} in each specific application of the NE-SCGLE theory. 

This synergic cooperation of equilibrium and non-equilibrium statistical thermodynamics of liquids is important because the advances just described represent only a sample of a wealth of scenarios that can emerge from further  applications and extensions of the NE-SCGLE theory. These excursions to more complex conditions will be greatly supported  if the rules of use of this new theoretical framework are identified and clarified in detail, and this is best done by means of specific examples. Thus, the second main aim of this work is to address this question with the application of the NE-SCGLE theory to a specific model fluid, namely, the hard-sphere plus square well (HSSW) fluid, focussing on the identification and definition of non-equilibrium phases and transitions, and on how they relate to the ``ordinary'' equilibrium phases. 

With these objectives in mind, our strategy will be to start in Section \ref{sect2} with a brief review of the  NE-SCGLE theory and its approximate thermodynamic input for the model system considered. There we also compare the ordinary equilibrium phase diagram with the SCGLE glass transition diagram and the NE-SCGLE non-equilibrium glass transition diagram. This is followed by the presentation in Section \ref{sect3} of the \emph{time-dependent} non-equilibrium phase diagram that results from the solution of the NE-SCGLE equations for the HSSW model liquid. We then conclude in Section \ref{sect4} with a brief summary of the main results of this work. There we also comment on the perspectives and potential impact of the concept of non-equilibrium time-dependent phase diagrams regarding the modeling and interpretation of actual experimental measurements of practical interest in material science and engineering.

\section{The NE-SCGLE theory and its application.}\label{sect2}

In this section we provide a summary of the Non-Equilibrium Self Consistent Generalized Langevin Equation (NE-SCGLE) theory, and determine the specific approximate free energy functional $\mathcal{F}[n,T]$ for the specific model system chosen to illustrate its concrete application. We then use this approximate $\mathcal{F}[n,T]$ to determine the corresponding equilibrium phase diagram, as well as the  glass transition diagram and its NE-SCGLE complement, whose discussion serves to introduce the concept of non-equilibrium time-dependent phase diagram in the following section.

\subsection{ The NE-SCGLE equations.}\label{subsect2.1}

The Non-Equilibrium Self Consistent Generalized Langevin Equation (NE-SCGLE) theory was proposed and explained in detail in Refs. \cite{nescgle1,nescgle3}. For details of its fundamental origin the interested reader is referred to  Ref. \cite{nescgle1}, and for brief review, to Section A of the supplemental material of Ref. \cite{nescgle8}, which highlights the main simplifying approximations leading to the lowest-order version of this theory, employed in this and in all previous applications.

Thus, let us only summarize here the NE-SCGLE theory in terms of the set of equations that describes the  non-equilibrium evolution of the structural and dynamical properties of a simple glass-forming liquid ($N$ identical spherical particles in a volume $V$ that interact through a radially-symmetric pair potential $u(r)$). It starts with the time evolution equation of the  non-equilibrium structure factor (SF) $S(k;t) \equiv \overline{ \delta n({\bf k};t) \, \delta n(-{\bf k};t)}$, where the over-line indicate the average over a (non-equilibrium) statistical ensemble, and where $\delta n({\bf k};t)$ is the Fourier transform of the fluctuations in the local particle number density  $n({\bf r};t)$. For a system that is instantaneously quenched at time $t=0$ from initial bulk density and temperature  $(n_i,T_i)$ to new final values $(n,T)$, such an equation reads for $t>0$ 
\begin{equation}
\frac{\partial S(k;t)}{\partial t} = -2k^2 D^0
b(t)n\mathcal{E}(k;n,T) \left[S(k;t)
-1/n\mathcal{E}(k;n,T)\right], \label{relsigmadif2pp}
\end{equation}
with  $D^0$ being the short-time self-diffusion coefficient, related by Einstein's relation  $D^0=k_BT/\zeta^0$ with the  corresponding short-time friction coefficient $\zeta^0$ (determined by its Stokes expression \cite{landaulifshitzfm} in colloidal liquids, or by the kinetic [or `Doppler' \cite{ornsteinuhlenbeck}] friction coefficient in the case of molecular liquids  \cite{atomic1}). In this equation $\mathcal{E}(k;n,T)$ is the Fourier transform (FT) of the second functional derivative $\mathcal{E}(r;n,T)$ of the Helmholtz free energy density-functional $\mathcal{F}[n;T]$ \cite{evans}, 
\begin{equation}
\mathcal{E}[\mid \mathbf{r}-\mathbf{r}'\mid ; n;T] \equiv  \left( \frac{\delta^2 \mathcal{F}[n;T]/k_BT}{\delta n(\mathbf{\mathbf{r}})\delta n(\mathbf{\mathbf{r}'})} \right) ,
\label{eqcondforsigmapp}
\end{equation}
evaluated at the uniform (bulk) density and temperature fields $n(\textbf{r},t)=n\equiv N/V$ and $T(\textbf{r},t)=T$, where $N,\ V$ are the total particle number and volume of the system and $T$ the final temperature of the quench. 

The  time-dependent mobility function $b(t)$ is the corresponding non-equilibrium value of the normalized (``long-time'') self-diffusion coefficient, 
\begin{equation}
b(t)\equiv \frac{ D_L(t)}{D_0}
\label{dlend0}. 
\end{equation}
This function couples the structural relaxation described by Eq. (\ref{relsigmadif2pp}) with the non-equilibrium relaxation of the dynamic properties of the fluid. Such coupling is established by the following exact expression for  $b(t)$,
\begin{equation}
b(t)= [1+\int_0^{\infty} d\tau\Delta{\zeta}^*(\tau; t)]^{-1},
\label{bdt}
\end{equation}
in terms of the $t$-evolving, $\tau$-dependent friction function $\Delta{\zeta}^*(\tau; t)$, for which the NE-SCGLE theory derives the following approximate expression \cite{nescgle1},
\begin{equation}
\begin{split}
  \Delta \zeta^* (\tau; t)= \frac{D_0}{24 \pi
^{3}n}
 \int d {\bf k}\ k^2 \left[\frac{ S(k;
t)-1}{S(k; t)}\right]^2  \\ \times F(k,\tau; t)F_S(k,\tau; t),
\end{split}
\label{dzdtquench}
\end{equation}
in terms of $S(k; t)$ and of the non-equilibrium  \emph{collective} intermediate scattering function (ISF) $F(k,\tau; t)\equiv N^{-1}  \overline{ \delta n(\mathbf{k},t+\tau) \delta n(-\mathbf{k},t)}$, where  $n(\mathbf{k},t)$ is the FT of the thermal fluctuations $\delta n(\mathbf{r},t)\equiv n(\mathbf{r},t)-n$ of the local number density  $n(\mathbf{r},t)$ at time $t$. The \emph{self} intermediate scattering function (self-ISF)  $F_S(k,\tau; t)$ is defined as $F_S(k,\tau; t)\equiv  \overline{ \exp \left[    i\mathbf{k}\cdot \Delta \mathbf{r}_T(t,\tau)    \right]}$, with $\Delta \mathbf{r}_T(t,\tau) \equiv  \left[ \mathbf{r}_T(t+\tau)-\mathbf{r}_T(t)\right]$ being the displacement of one particle considered as a  tracer. In both cases, the over-line indicates  an average over a corresponding non-equilibrium statistical ensemble.

These equations are complemented by the  memory-function equations for  $F(k,\tau; t)$ and $F_S(k,\tau; t)$, written approximately, in terms of the Laplace transforms (LT) $F(k,z; t)$ and $F_S(k,z; t)$, as
\begin{gather}\label{fluctquench}
 F(k,z; t) = \frac{S(k; t)}{z+\frac{k^2D^0 S^{-1}(k;
t)}{1+\lambda (k)\ \Delta \zeta^*(z; t)}},
\end{gather}
and
\begin{gather}\label{fluctsquench}
 F_S(k,z; t) = \frac{1}{z+\frac{k^2D^0 }{1+\lambda (k)\ \Delta
\zeta^*(z; t)}},
\end{gather}
where
\begin{equation}
\lambda (k)\equiv 1/[1+( k/k_{c}) ^{2}]
\label{lambdadk}
\end{equation}
is an ``interpolating function" \cite{todos1}, with $k_c$ being an empirically determined parameter. In the present work we use $k_c= 1.305(2\pi)/\sigma$, with $\sigma$ being the hard-core particle diameter of our HSSW model fluid, which guarantees that the hard-sphere liquid will have its dynamic arrest transition at a volume fraction $\phi_c=0.582$, in agreement with simulations  \cite{gabriel}. 

Eq. (\ref{relsigmadif2pp}), together  with Eqs. (\ref{bdt})-(\ref{lambdadk}), constitute a closed system of equations that summarizes the  NE-SCGLE theory. If the Helmholtz free energy density functional $F=\mathcal{F}[n;T]$ (or at least its second functional derivative $\mathcal{E}[r ; n;T]$) has been determined, then the solution of these equations yield predictions for $S(k;t)$ and for the dynamic properties  $b(t)$, $\Delta{\zeta}^*(\tau; t)$, $F(k,\tau; t)$, and $F_S(k,\tau; t)$.

\begin{figure*}
\begin{center}
\includegraphics[scale=.43]{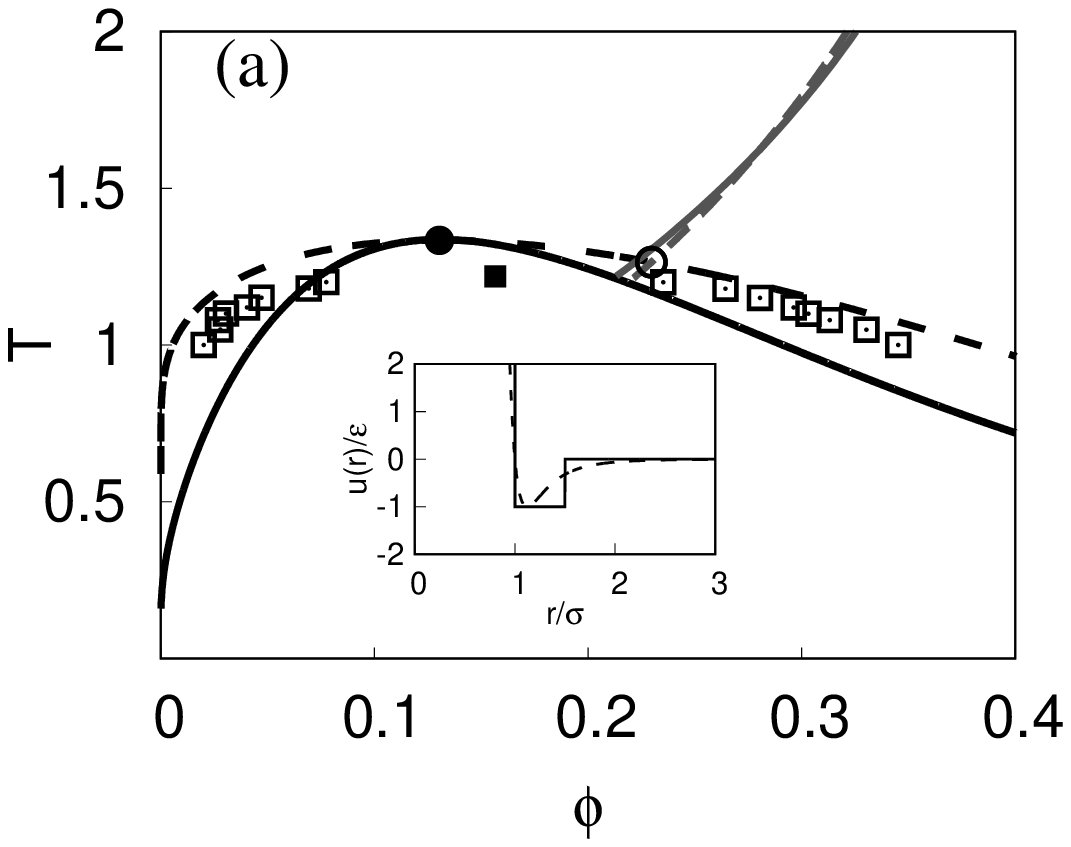}
\includegraphics[scale=.44]{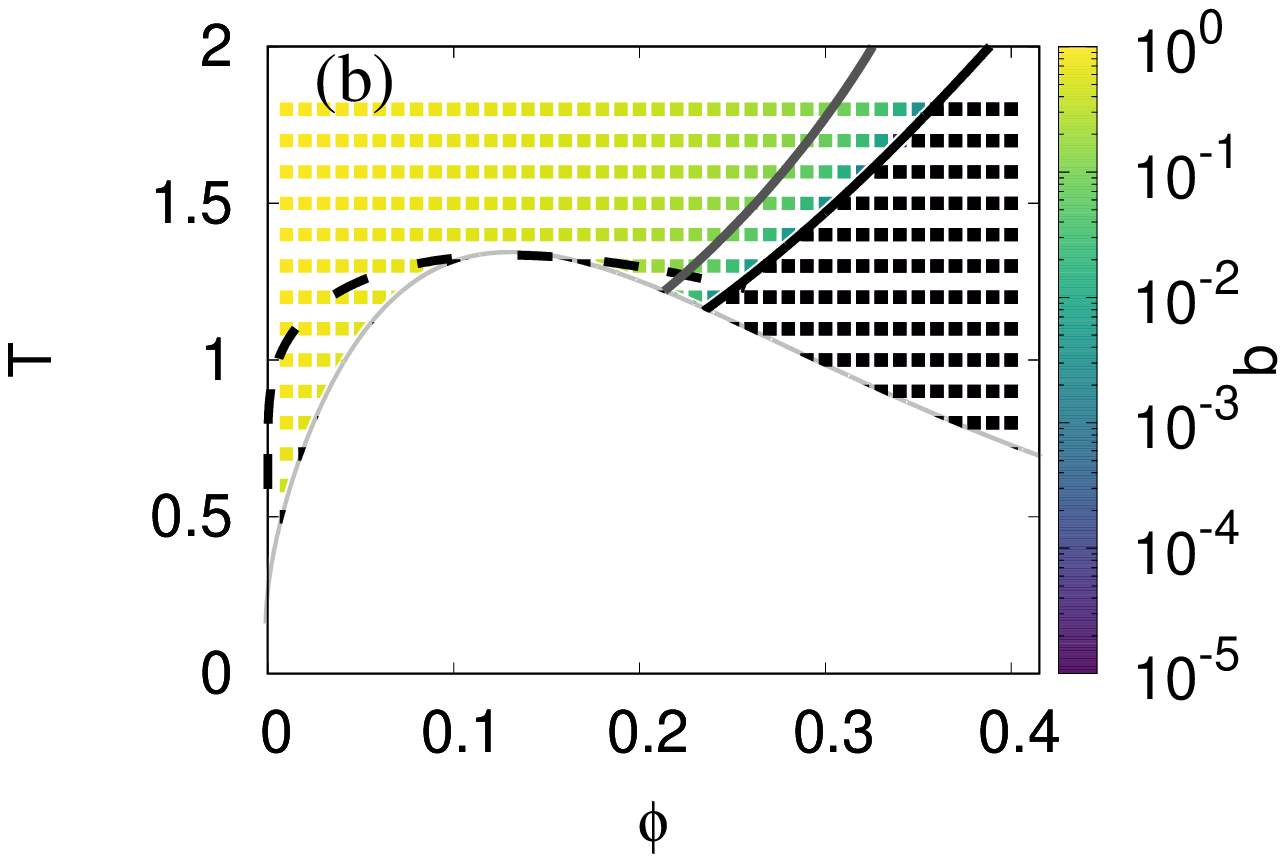}
\includegraphics[scale=.44]{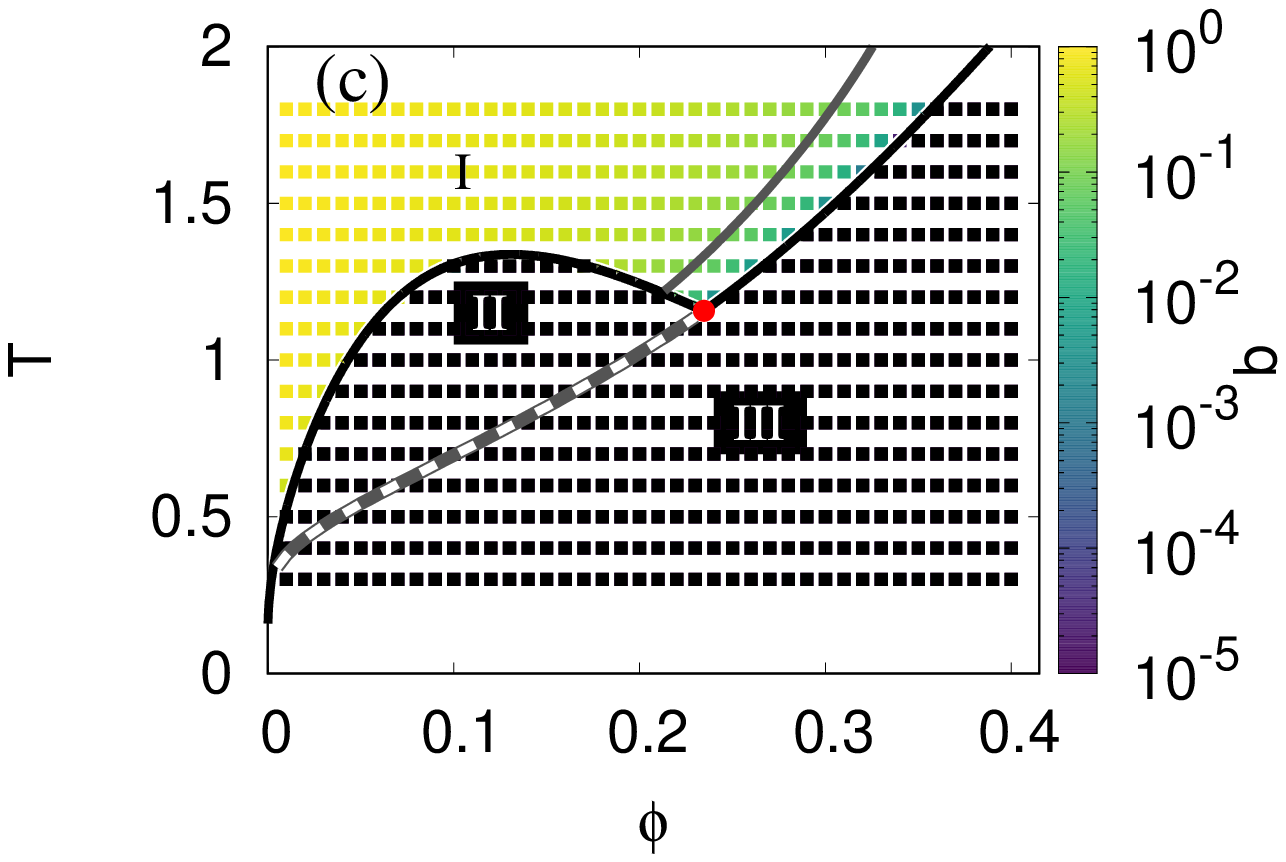} 
\caption{(a) Equilibrium phase diagram of the HSSW model for $\lambda=1.5$. The theoretical predictions of the spinodal and binodal gas-liquid coexistence are respectively represented by the black solid and black dashed lines, with the black circle being the critical point. Simulation data of the binodal and critical point are represented by the empty and full black squares (obtained from Ref. \cite{vega1992}). The gray solid and dashed lines represent the freezing lines, respectively obtained through L\"owen's and Hansen-Verlet freezing criteria. The inset shows a comparison between the 12-6 Lennard-Jones interaction potential (dashed line) and the HSSW interaction potential (solid line). (b) Glass transition diagram, obtained through the SCGLE (equilibrium) theory. The black solid line represents the fluid-glass transition line. The predicted equilibrium mobility (colored region) is represented through a color palette for points outside the spinodal (clear gray solid line enclosing the lower white region). (c) Non-equilibrium glass transition diagram obtained through the NE-SCGLE theory. The black solid line is the boundary between arrested (regions II and III) and fluid states (region I). The black dashed line represents the predicted gel-glass transition line. The red circle point represents the bifurcation point, where the equilibrium glass transition loci intersect the spinodal. Glass-like arrested states are expected to exist within region III, while gel-like arrested states in region II. Each mobility color corresponds to the asymptotic value of the mobility $\lim_{t\to\infty}b(t)$ for instantaneous isochoric quenches, where the black has been reserved for $\lim_{t\to\infty}b(t)=0$.}
\label{fig1}
\end{center}
\end{figure*}

\subsection{Model system and equilibrium phase diagram.}\label{subsect2.2}

The first step in each application of the  NE-SCGLE theory is the determination of its essential thermodynamic input, namely, the Helmholtz free energy density functional $F=\mathcal{F}[n;T]$ of the system of interest, or directly its second functional derivative $\mathcal{E}(r;n,T)$ in Eq.(\ref{eqcondforsigmapp}). Statistical thermodynamics  has produced a variety of approximate solutions \cite{mcquarrie,hansen}. Refs. \cite{foffidawson2002,sperl1}, for example, provide  detailed studies of simple model fluids consisting of particles interacting through hard-core plus a short-range attractive (Yukawa or square-well) potentials. Using simple but accurate approximations for $F=\mathcal{F}[n;T]$ they can generate the  equilibrium phase diagram. 

%and through mode-coupling theory, they determine the dynamic arrest diagram, i.e., the boundary between the equilibrium supercooled fluid and the non-equilibrium glass phases. Thus, from the same approximate free energy, the \emph{equilibrium} and the (MCT-like) \emph{non-equilibrium}  phase diagrams are derived, so that both can be discussed in reference to each other. 

%In Fig. \ref{fig1} 

%Here we carry out an analogous exercise, but with the intention to compare with the theoretical concept of \emph{time-dependent non-equilibrium} phase diagram developed below. For this, among the model systems employed in similar studies \cite{fabian,bergenholtzfuchs,foffidawson2002}, let us adopt 

In Fig. \ref{fig1}(a) we carry out an analogous exercise
using the (Brownian) hard-sphere plus square well (HSSW) model, formed by  $N$ spherical particles in a volume $V$ interacting through the pair potential 
\begin{equation}
u(r)=
\begin{cases}
\infty, & (r/\sigma)< 1; \\
-\epsilon, & 1\le (r/\sigma) \le \lambda; \\
0, &  (r/\sigma) > \lambda.
\end{cases}
\label{hssw}
\end{equation}
In the inset of Fig. \ref{fig1}(a) we plot this pair potential with $\lambda=1.5$ (solid line). As a reference, we also plot the Lennard-Jones potential $u(r)= 4\epsilon [(\sigma/r)^{12}-(\sigma/r)^{6}]$ (dashed line). Both model systems lead to essentially the same physics, but the HSSW is analytically simpler.

The state space of our HSSW model is spanned by the dimensionless number density $[n\sigma^3]$ and temperature $[k_BT/\epsilon]$ (with $k_B$ being Boltzmann's constant).  From now on, we shall use $\sigma$, $[\sigma^2/D_0]$, and $\epsilon$ as the units of length, time, and energy, and denote $[n\sigma^3]$ and $[k_BT/\epsilon]$ simply as $n$ and $T$, so that the hard-sphere volume fraction is $\phi\equiv \pi n/6$. The dimensionless time $[D_0t/\sigma^2]$ will be denoted simply as $t$. At each state point $(n,T)$  the Helmholtz free energy $\mathcal{F}[n,T]$ must be defined, for which we rely on the so-called modified mean field (MMF) approximation  \cite{groh,teixeira,frodldietrich,tavares}, defined in detail in Appendix \ref{appendix1}. By construction, approximations such as  this are meant to describe only spatially homogeneous and isotropic (i.e., gas and liquid) equilibrium phases, thus filtering out crystalline solid phases, very much as polydispersity does in colloidal systems \cite{hspolydispersity}. Thus, it allows the determination of the  equilibrium phase diagram of the HSSW fluid  illustrated in Fig. \ref{fig1}(a). It consists of the binodal and spinodal lines of the gas-liquid transition, and of the  (dashed) line of constant height of the main peak of $S^{eq}(k;n,T)= 3.1$ (the long-time equilibrium stationary  solution $S(k;t\to\infty)= S^{eq}(k;n,T)\equiv 1/ n\mathcal{E}(k;n,T)$ of  Eq. (\ref{relsigmadif2pp})); according to Hansen and Verlet \cite{hansenverlet}, this is  a proxy of the freezing line. Fig. \ref{fig1}(a) also includes, for reference, the exact simulation data of Ref. \cite{vega1992} for the binodal curve.

This conventional equilibrium phase diagram serves as a reference to define what we shall call \emph{non-equilibrium phase diagrams}, defined not by the usual equilibrium thermodynamic criteria \cite{callen,mcquarrie}, but by kinetic or dynamic order parameters. The first kind of such non-equilibrium phase diagrams was theoretically defined in the framework of MCT. It consists of the boundary between the ergodic region of state space $(n,T)$, where the system will be able to reach the corresponding thermodynamic equilibrium state, and the non-ergodic region, where it will be trapped in  kinetically-arrested states. The resulting non-equilibrium phase diagrams are also referred to as \emph{glass transition}\cite{sperl1} (or \emph{dynamic arrest\cite{pedroatractivos}) diagrams}.

\subsection{Glass transition diagram.}\label{subsect2.3}

Let us first contrast the equilibrium phase diagram of Fig. \ref{fig1}(a) with the MCT concept of \emph{glass transition diagram} obtained from the MCT equations (e.g., Eqs. (1) and (3) of Ref. \cite{sperl1}), more specifically, from  the solution of the asymptotic form of these equations (i.e., Eq. (2) of Ref. \cite{sperl1}) for $f(k;n,T)\equiv \lim_{\tau\to\infty} F^{eq}(k,\tau;n,T)$. This order parameter vanishes at equilibrium and differs from 0 at non-ergodic states. The results for  $f(k;n,T)$ thus partitions the state space $(n,T)$ into the ergodic and  the non-ergodic regions, with the ideal glass transition line as the boundary between them. A detailed and careful report of the resulting glass transition diagram for the HSSW system is provided by Sperl in Ref. \cite{sperl1}, although only for very short-ranged attractions $(\lambda-1)\ll 0.5$ and very high volume fractions ($\phi>0.5$), where the reentrance from repulsive glass-to-fluid-to-attractive glass was predicted and observed\cite{cardinaux,gibaud,luetalnature,Gao,foffi2014}. To conform to previous work \cite{nescgle5,nescgle7,nescgle8}, however, here we shall restrict ourselves to $\lambda=1.5$,  which is realistic for atomic liquids (and for some particular colloidal systems \cite{royalturci,espinosafrisken}), leaving the analysis of shorter ranges for a separate study.

The main features of the resulting glass-transition (or dynamic arrest) diagram are presented in Fig. \ref{fig1}(b). This non-equilibrium phase diagram was obtained, however, not from the MCT equations, but following the procedure first explained in Ref. \cite{todos1}, based on the  \emph{equilibrium} SCGLE theory of dynamic arrest (see a brief summary in section 1 of Appendix B). We refer to this diagram as glass transition diagram because the SCGLE theory is  analogous to MCT in most respects, including the qualitative features of the long-time asymptotic scenario summarized in the figure. For example, neither of these equilibrium theories is applicable when the system is quenched inside the spinodal region (white region in the figure), where $S^{eq}(k;n,T)$ does not exist.

More important, both theories predict two kinetically-complementary regions, namely, the (colored) region of equilibrium ergodic states, where  the equilibrium particle mobility  $b^{eq}(\phi,T)$ is non-zero, and the region where $b^{eq}(\phi,T)$ vanishes (colored in black), and the system is predicted to get trapped in a dynamically-arrested state. The boundary $T=T_c(\phi)$ is the dynamic arrest line (or ``ideal glass transition'' line), defined as the limiting iso-diffusivity line $b^{eq}(\phi,T)= 0$. Other iso-diffusivity lines are defined by the condition of constant $b^{eq}(\phi,T)= b_0$, with $b_0>0$.  Fig. \ref{fig1}(b) includes the case $b_0=0.1$, another proxy of the freezing line according to L\"owen's dynamical criterion for freezing \cite{lowen}. Notice, however, that the SCGLE theory does not distinguish stable from metastable equilibrium states, and treats them on the same footing. Thus, the freezing and the binodal lines are only included for reference in \ref{fig1}(b), since they have no real
dynamic significance. 

\subsection{NE-SCGLE non-equilibrium glass transition diagram.}\label{subsect2.4}

Since the equilibrium SCGLE theory is an asymptotic stationary limit of the NE-SCGLE theory, one wonders if both theories will yield the same  glass transition diagram. This question was originally posed and answered in detail in Ref. \cite{nescgle5},  resulting in a practical protocol  briefly explained in section 2 of Appendix B. The result of its application to our HSSW model is summarized in the NE-SCGLE non-equilibrium glass transition diagram in Fig. \ref{fig1}(c), whose predictions regarding the equilibration (colored) and the dynamically arrested (black)  regions above the spinodal curve coincide with those of the equilibrium theory. In contrast with the latter, however,  the NE-SCGLE theory is perfectly applicable at and below the spinodal curve, yielding the scenario illustrated in  Fig. \ref{fig1}(c), in which the spinodal curve, besides being the threshold of the thermodynamic stability of homogeneous states, is also predicted to be the borderline between the regions of ergodic and non-ergodic homogeneous states. In addition, the high-density liquid-glass transition line, whose high-temperature limit corresponds to the well-known hard-sphere glass transition,  at lower temperature intersects the spinodal curve and continues inside the spinodal region as a gel-glass transition line. 

The equilibrium region of state space, identified by the condition that $b^{eq}(\phi,T) >0$, is bounded from below by this dynamic-arrest line,  represented in Fig. \ref{fig1}(c) by  the solid line, one segment of which coincides with the spinodal curve, and the other being the ideal liquid-glass transition.  At and below this composed curve the asymptotic square localization length $\gamma_a(\phi,T)$ (see Appendix A.4) is finite, while it is infinite above. As explained in Ref. \cite{nescgle5},  the difference between these two segments is that  $\gamma_a(\phi,T)$ is a continuous (``type A'') function of $T$ along the spinodal, but is discontinuous (``type B'') along the liquid-glass line. This ``type B'' discontinuity, however, is also predicted to continue as a gel-glass transition along the short-dashed line \cite{nescgle5}.  Strictly speaking, however, these sharp transitions will never be observed in practice, since they would only occur at $t=\infty$.  In any real experiment one will always observe a finite-time prediction, such as those represented in Fig.  \ref{fig3}. Depending on the conditions, however, one might observe an earlier or a later stage of the aging process.

\begin{figure*}[ht!]
\centering
    \includegraphics[width=.3\textwidth]{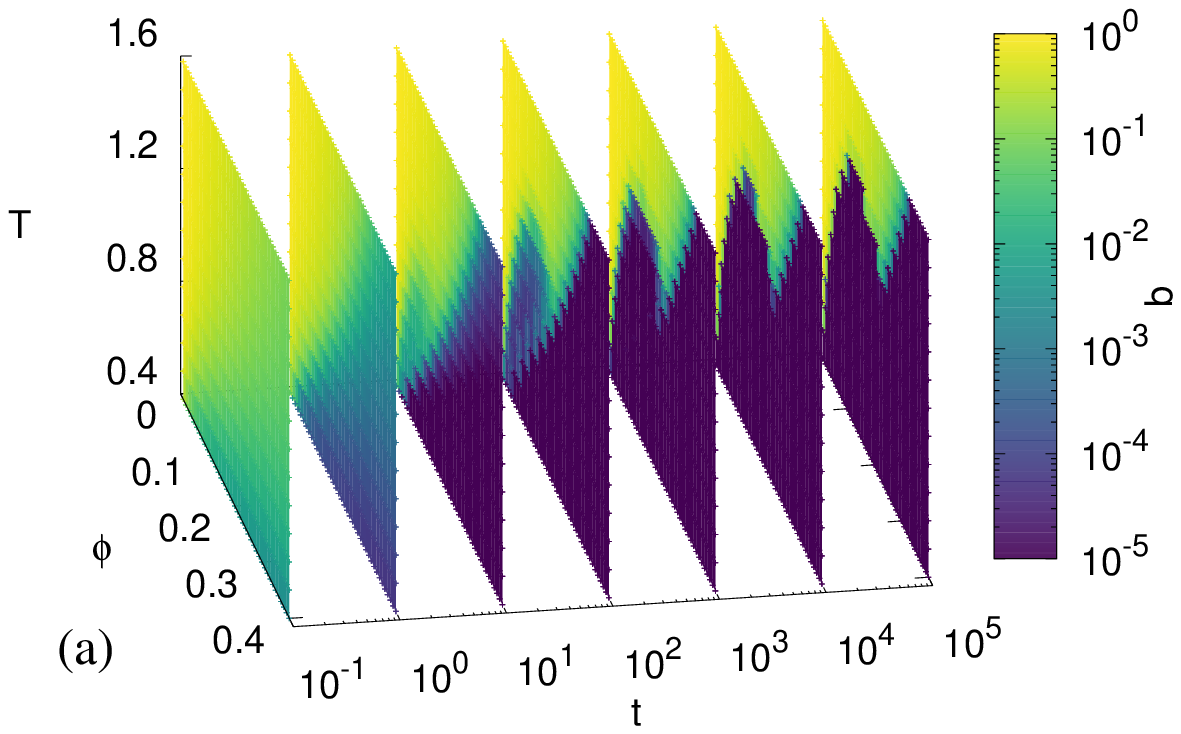}
    \includegraphics[width=.3\textwidth]{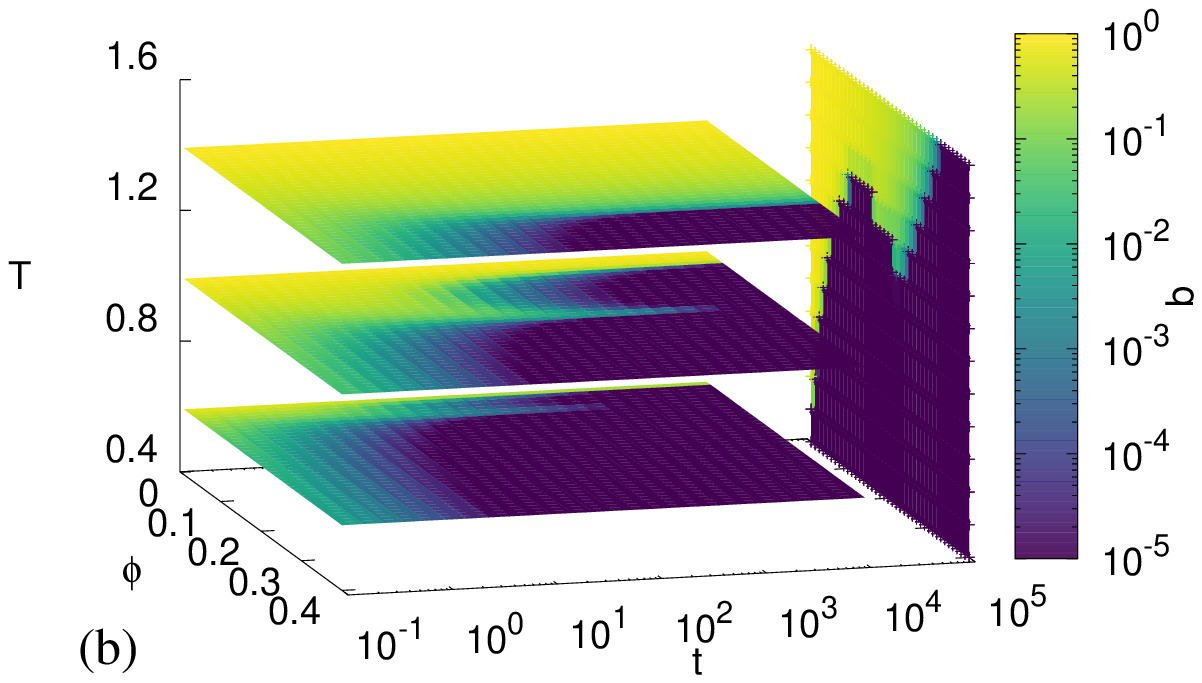}   
    \includegraphics[width=.3\textwidth]{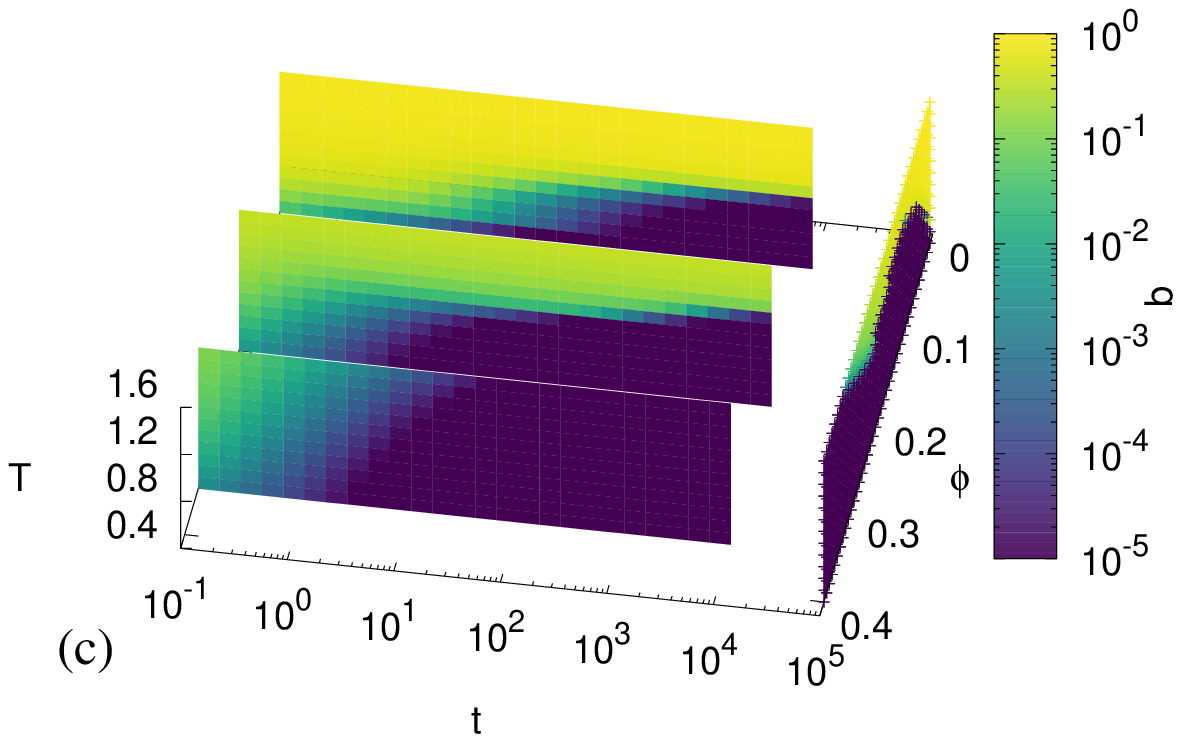}   
    \caption{Different planes of $b(t;T;\phi)$. (a) $\phi$-$T$ planes at constant $t=10^{-1}$, $10^{0}$, $10^{1}$, $10^{2}$, $10^{3}$, $10^{4}$ and $10^{5}$. (b) $\phi$-$t$ planes at constant $T=0.6$, $1.0$ and $1.4$, along with the $\phi$-$T$ plane at $t=10^{5}$. (c) $t$-$T$ planes at constant $\phi=0.05$, $0.20$ and $0.35$, along with the $\phi$-$T$ plane at $t=10^{5}$.}
\label{fig2}
\end{figure*}

\subsection{Physical meaning and the need of a kinetic perspective.}\label{subsect2.5}

There are some subtle aspects in the physical interpretation of the three diagrams in Fig. \ref{fig1} and the relationship between them, that must be mentioned. Of course, this does not refer to the equilibrium (colored) region, in which the three diagrams agree completely. The only difference in this region is that the free-energy determines only the thermodynamic state functions (pressure, energy, etc.), while the dynamic theories (MCT or SCGLE) determine also  the main \emph{dynamic} state functions (mobility, relaxation time, viscosity, etc.), illustrated by the color scale in panels (b) and (c), which displays $b^{eq}(\phi,T)$. 

The subtleties and conceptual difficulties arise, however, in the understanding of the dynamic arrest condition $b^{eq}(\phi,T)=0$ represented by the dark region of Fig. \ref{fig1}(c) and by the limiting dynamic arrest (solid) line. The reason is that this condition implies infinite relaxation times, whose observation requires strictly infinite experimental times. Thus, in a real (or simulated) experiment, if we quench a system at a state point $(\phi,T)$ in the dark region, we should be prepared to measure some finite non-equilibrium value $b(t;\phi,T)$, corresponding to the finite experimental waiting time $t$, but never in practice the asymptotic value $b^{eq}(\phi,T)=0$. Unfortunately, because of their equilibrium nature, neither MCT or the equilibrium SCGLE theory are capable of predicting $b(t;\phi,T)$ and the other relevant dynamic properties for practical (i.e.,  finite) experimental times $t$. In contrast, this is precisely the main contribution of the NE-SCGLE theory, which thus provides the kinetic perspective needed to better connect theory with real experiments or simulations. The results of the following section illustrate this important contribution.

Let us clarify again that the present version of the NE-SCGLE theory does not yet contemplate a kinetic pathway to crystallization or to non-uniform macroscopic gas-liquid phase separation. In real experiments, however, the formation of glasses competes with crystallization, and the predicted dynamically arrested spinodal decomposition competes with the possibility that the system macroscopically phase-separates to reach its stable heterogeneous equilibrium state. Possible experimental filters, however, may be thought of. For example, for crystal formation, a possible filter may be the introduction of polydispersity  \cite{hspolydispersity}, whereas the spontaneous (micro-) phase separation resulting from arrested spinodal decomposition, may impede the formation of infinite crystalline structures. As a result, the scenario represented by Figs. \ref{fig1}(b) and (c) may eventually be enriched by  the inclusion of these or other experimentally relevant additional effects. However, for the time being their absence will make the discussion somewhat simpler and precise.

\section{Time-dependent Non-equilibrium phase diagram.}\label{sect3}

Let us thus present the main results of this work, which derive from the solution of the  NE-SCGLE equations (Eq. (\ref{relsigmadif2pp})  with Eqs. (\ref{bdt})-(\ref{lambdadk})) for the non-equilibrium structural and dynamical properties  of our HSSW model liquid. Let us mention that the predictions for the evolution of some of these properties in \emph{individual} temperature quenches, have already been successfully compared with concrete simulations and experiments \cite{nescgle5,nescgle7,nescgle8}. In this work, in contrast, we are interested in understanding the scenario that emerges when we consider not a single quench, but an ensemble of many simultaneous such quenches. In addition, rather than discussing the evolution of several properties, here we shall focus on only one physically meaningful property, namely, the non-equilibrium mobility function $b(t)$.

To have a concrete idea in mind, let us imagine that we prepare an ensemble of samples, assumed originally (for times $t<0$) in thermodynamic  equilibrium at volume fraction $\phi$ (for a set of volume fractions) and initial temperature $T_i$, which for concreteness we set as $T_i =\infty$, so that the initial state is equivalent to an equilibrium hard-sphere liquid at volume fraction $\phi$. We then assume that the attractive interactions are suddenly turned on, i.e., that at time $t=0$ these initial equilibrium systems are  instantaneously and  isochorically quenched to a finite final temperature $T$, kept fixed afterwards ($t>0$).  We then solve  the NE-SCGLE equations  with the aim of describing the \emph{time-dependent non-equilibrium phase diagram} that emerges  from considering the ensemble of simultaneous such quenches that differ in  their final state point $(\phi,T)$.  In what follows  we shall illustrate this idea with the explicit results for the set of final state points indicated by the matrix of colored symbols in Fig. \ref{fig1}(c).

For each individual quench, $b(t)$ will depend on the initial and final state points $(\phi_i,T_i)$ and $(\phi,T)$. For simplicity, however, we have chosen each quench to be isochoric, $\phi_i=\phi$, and as for the initial temperature, we choose $T_i=\infty$. Thus, we shall consider $b(t)$ as a function only of the final state point $(\phi,T)$, so as to analyze  $b(t;\phi,T)$ as a function of only these three independent arguments. Since the asymptotic long-time limit $b_a(\phi,T)\equiv b(t\to\infty;\phi,T)$ served as the dynamic order parameter leading to the non-equilibrium glass transition diagram in Fig. \ref{fig1}(c), this function must embody the finite-time version of our dynamic arrest diagram. The simplest representation of this function is provided by Fig. \ref{fig2}, which uses a color scale to indicate the value of the function $b(t;\phi,T)$ in the three-dimensional parameter space $(t;\phi,T)$. Figs. \ref{fig2}(a), (b), and (c) visualize the cross sections of this function along planes of constant $t$, $T$, and $\phi$, respectively. Each of these visual perspectives, however, offers a different perspective of the function  $b(t;\phi,T)$. As we now see, each perspective highlights a different physical message and meaning. 

The most complete of them is presented in Fig. \ref{fig2}(a), which displays a sequence of snapshots in the plane $(\phi,T)$ corresponding to a sequence of waiting times $t$. The second,  in Fig. \ref{fig2}(b),  displays the evolution of $b(t;\phi,T)$ for subsets of quenches with the same final temperature $T$. Finally, Fig. \ref{fig2}(c) presents the evolution of subsets of quenches with the same volume fraction. Let us now analyze each of these sequences.

\begin{figure*}
	\centering
    \includegraphics[width=.32\textwidth]{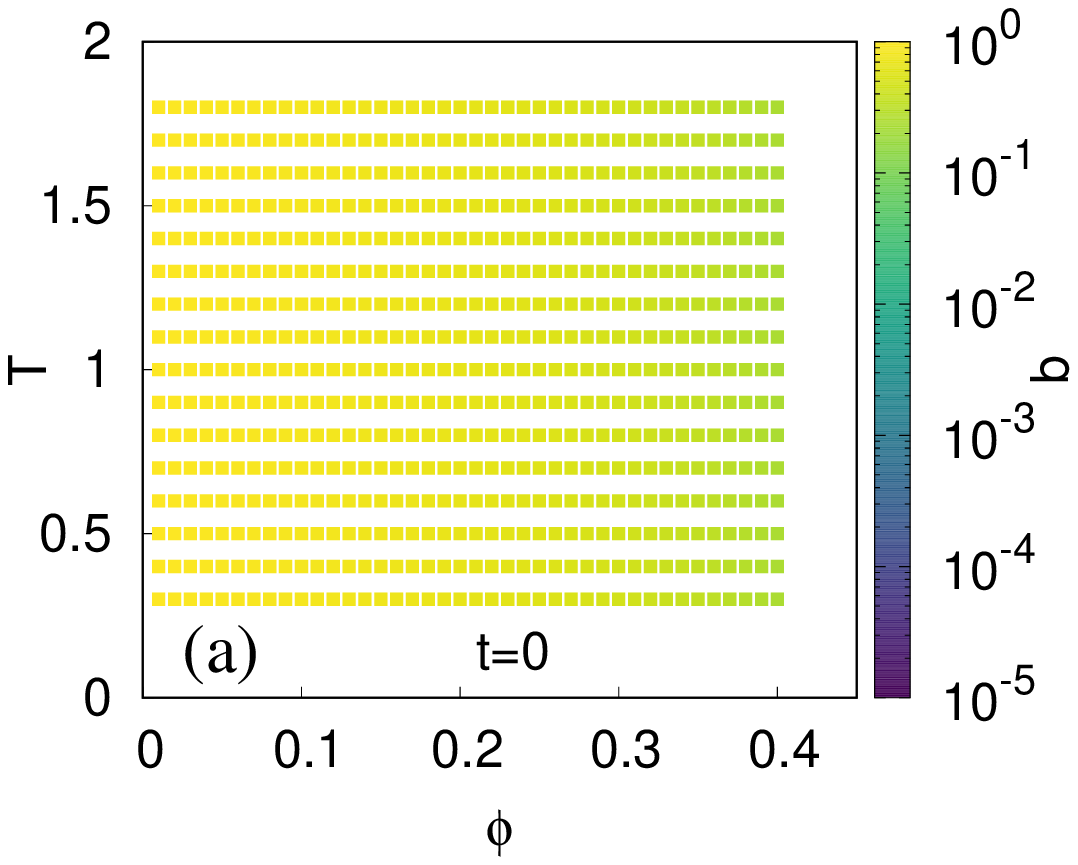}
    \includegraphics[width=.32\textwidth]{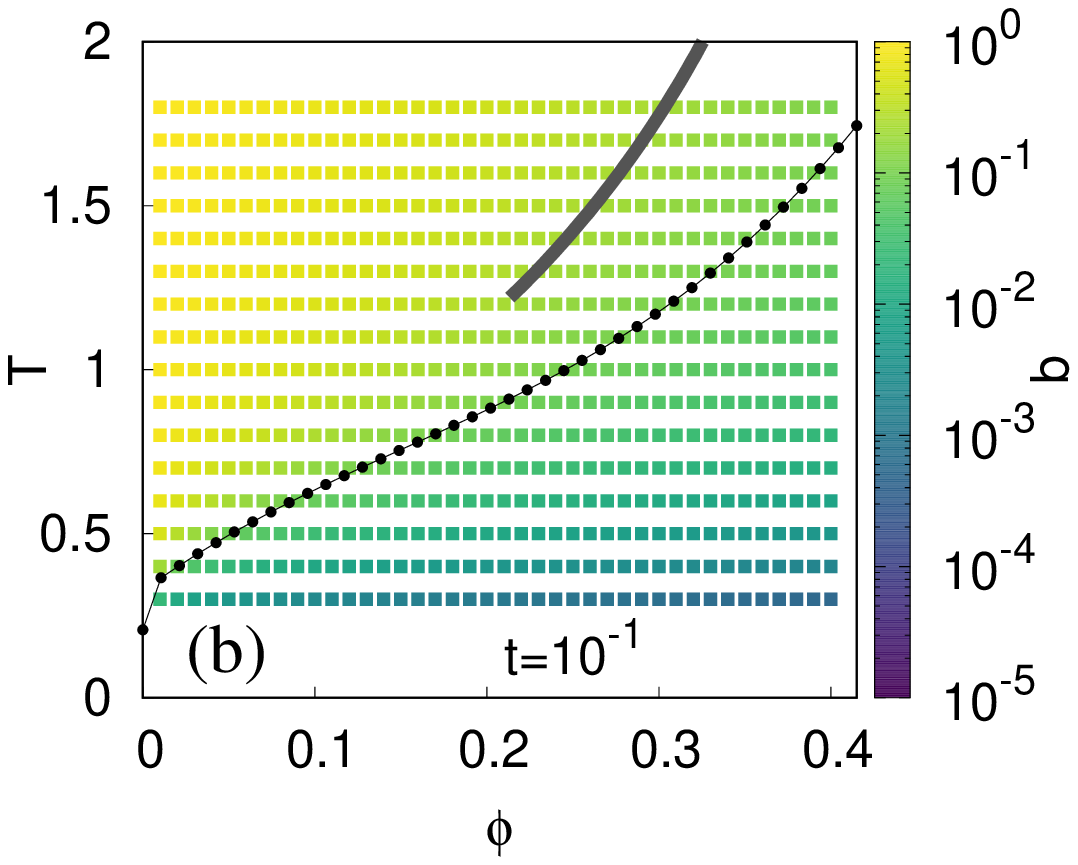}
	\includegraphics[width=.32\textwidth]{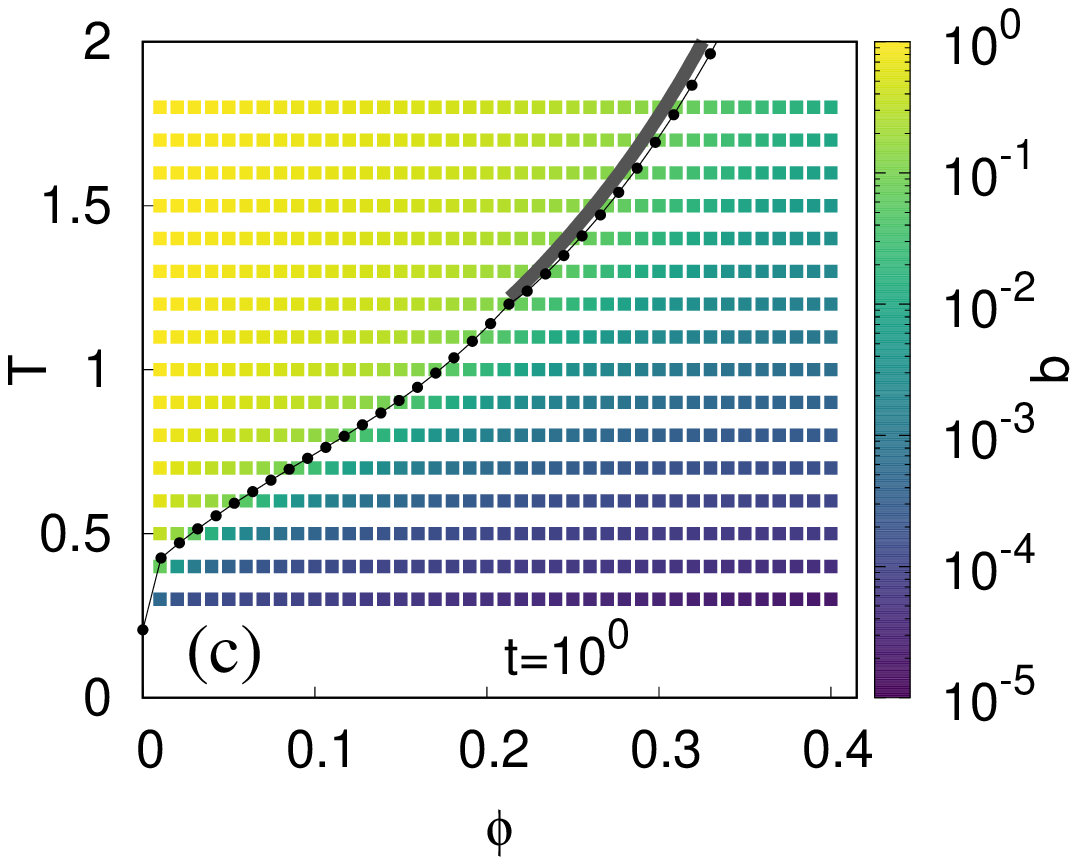}
	\includegraphics[width=.32\textwidth]{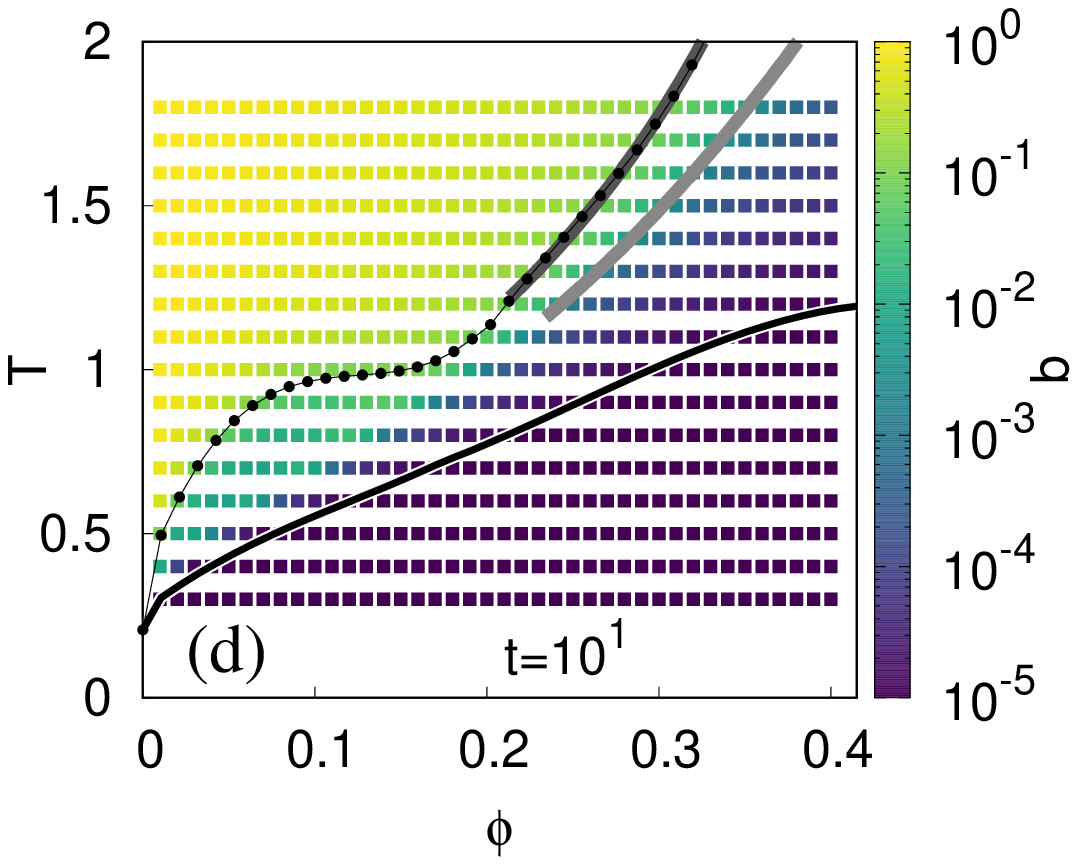}
    \includegraphics[width=.32\textwidth]{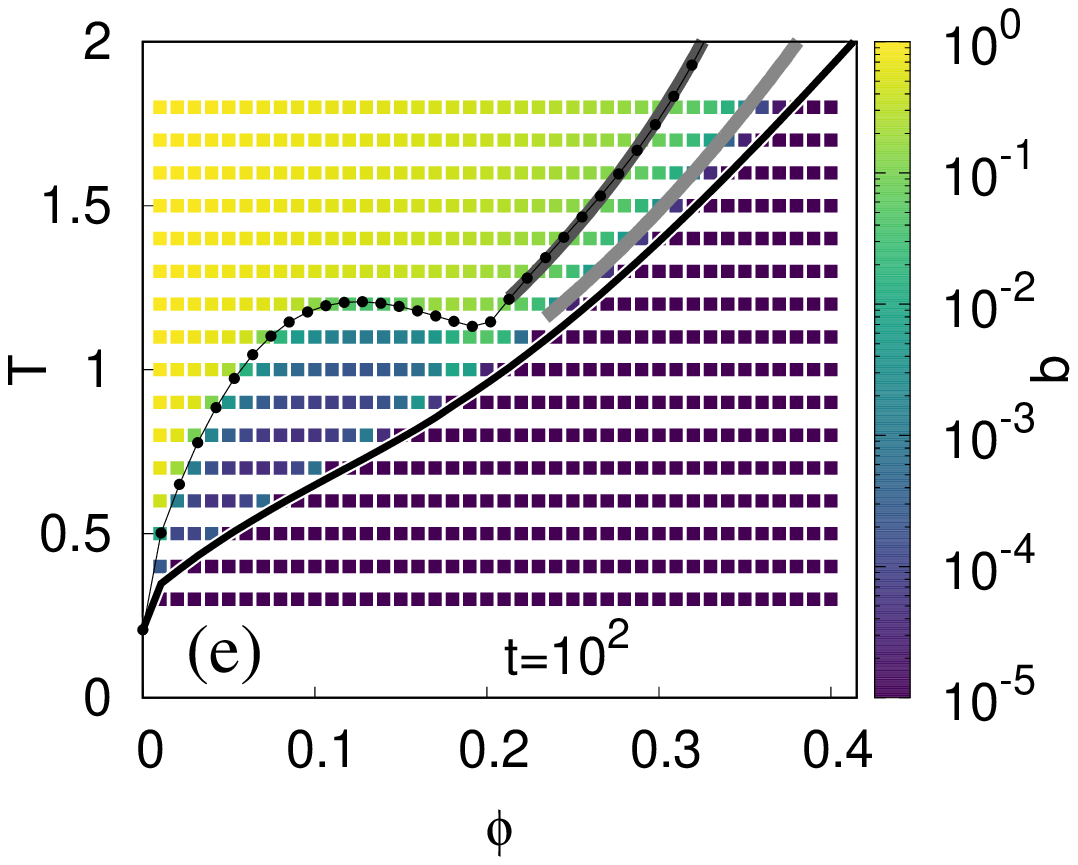}
    \includegraphics[width=.32\textwidth]{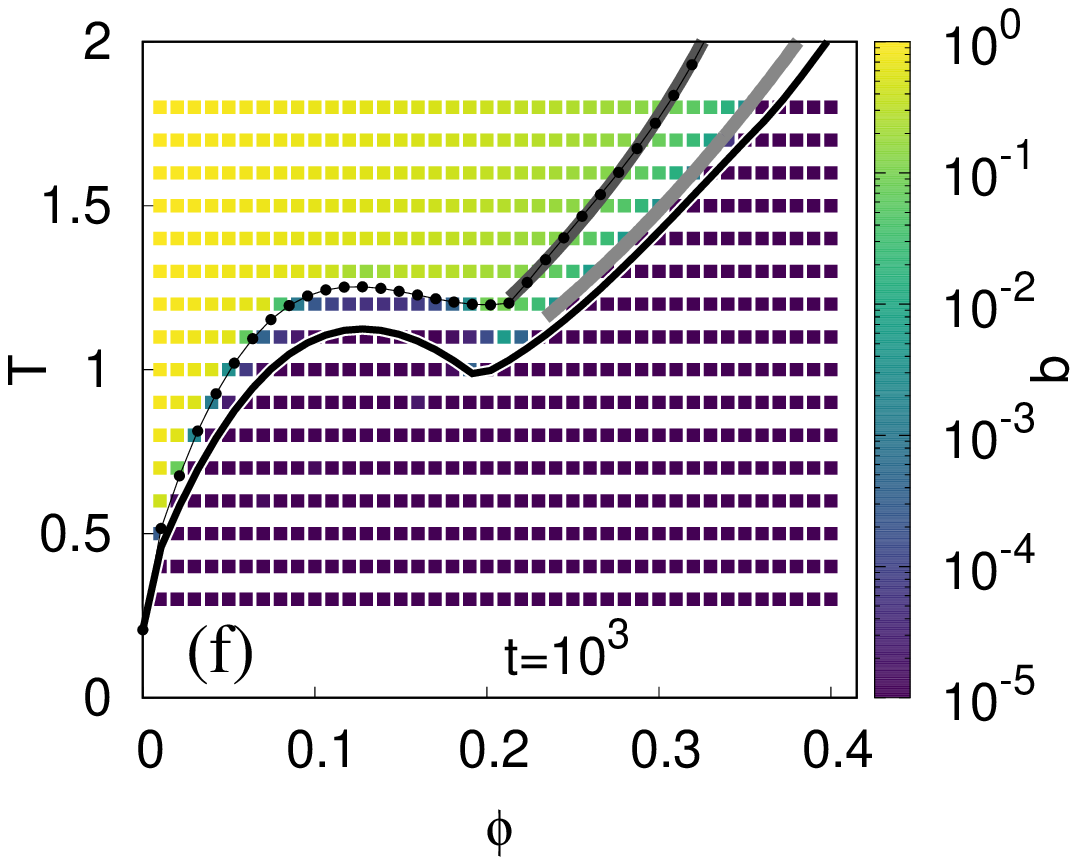}
  \caption{Evolution of the $\phi$-$T$ planes of $b(t;\phi,T)$ for an instantaneously-quenched HSSW model fluid at $t=0$ (a), $10^{-1}$ (b), $10^{0}$ (c), $10^{1}$ (d), $10^{2}$ (e) and $10^{3}$ (f). The black dotted (b)-(f) and solid (d)-(f)  lines corresponds to the iso-diffusivity (contour) lines  $b(t;\phi,T)=10^{-1}$ and $10^{-5}$ respectively. The thick solid dark gray (b)-(f) and clear gray (d)-(f) lines corresponds to the asymptotic equilibrium iso-diffusivity lines $b^{eq}(\phi,T)=10^{-1}$ and $10^{-5}$ respectively.}

\label{fig3}
\end{figure*}

\subsection{Snapshots in the plane $(\phi,T)$ at a sequence of waiting times.}\label{subsect3.1}

Let us start by discussing in more detail the sequence of snapshots presented in Fig. \ref{fig2}(a). This is done in Fig. \ref{fig3}, using  the same chromatic code to visualize the non-equilibrium evolution of the magnitude of $b(t;\phi,T)$ as a function of $\phi$ and $T$ at fixed  waiting times. The six snapshots correspond to the waiting times $t=\ 0, \ 10^{-1}, \ 10^{0}, \ 10^{1}, \ 10^{2}$, and $10^{3}$, which illustrate the early and intermediate stages of the process.  The long-time scenario is established, within the resolution of the figures, already at $t=\ 10^{5}$, whose snapshot is already indistinguishable from the asymptotic limit $t=\infty$ shown in Fig. \ref{fig1}(c). 

The first feature to observe is, of course, the monotonic decrease of $b(t;\phi,T)$ with $t$, indicating the irreversible slowing down of the dynamics. This process  is visualized  as a progressive darkening of each state point as $t$ evolves, although its speed clearly depends on the state point $(\phi,T)$. Thus, the coloring  is strongly inhomogeneous, with  $b(t;\phi,T)$ reaching a given lower threshold  faster in some regions than in others. A useful manner to enhance the visualization of the evolution of $b(t;\phi,T)$ is through  the concept of iso-diffusivity lines, first employed in equilibrium simulations by Zaccarelli et al. \cite{zaccarellifoffidawson2002} Extended to non-equilibrium, this concept refers to the loci in the plane  $(\phi,T)$ of the points where the mobility has the same value at a given waiting time $t$, say  $b(t;\phi,T)=b_0$. Figs.  \ref{fig3}(a)-(e) illustrate this concept with the iso-$b$  lines  corresponding to $b_0= 10^{-1}$ (black dotted line)  and $ 10^{-5}$ (black solid line). At $t=0$ none of them appear within the $(\phi,T)$-window  illustrated in the figure. The reason is that the initial condition $b(t=0;\phi,T)=b_{HS}(\phi)$ implies that all the iso-diffusivity lines are initially vertical lines, and the iso-$b$ lines corresponding to $10^{-1}$ and $ 10^{-5}$ are the isochores $\phi=$ 0.49 and 0.58, which  lie to the right of this window.

After some time, however, the mobility of some points within the illustrated window decreases below $b_0$. This occurs rather fast for $b_0= 10^{-1}$, as illustrated in Figs.  \ref{fig3}(b) and (c), which show that the dotted iso-$b$ line has moved to the center of this window, to quickly coincide with its long-time limiting line $b(t\to\infty;\phi,T)=b^{eq}(\phi,T) = 0.1$. According to the dynamic criterion of freezing of Ref. \cite{lowen}, this limiting iso-$b$ line lies close to the equilibrium freezing line $T=T_f(\phi)$. Here, however, it separates the stable from the metastable liquid regions.  Notice that in the early stage illustrated in Figs.  \ref{fig3}(a)-(c), the  iso-$b$ line $b(t;\phi,T) = 0.1$ appears as a continuous and monotonically increasing function of $\phi$. 

This early stage ($t \lesssim 1$) can only involve fast diffusion-limited clustering of nearest-neighbor particles, leading to the  final scenario in Fig.  \ref{fig3}(c). This final scenario, however, is just the beginning of the ensuing story (times $t>>1$), involving much slower collective restructuring processes. In this longer second stage  other striking and relevant features develop, which are illustrated in Figs.  \ref{fig3}(d)-(f). The first is the  kinetic difference between the high-$T$ segment of the iso-$b$ line $b(t;\phi,T)= 0.1$, which for $t>10$ remains stationary, and its lower  segment, which starts a slow invasion of the region where the equilibrium gas-liquid coexistence should be located.

\begin{figure*}
  \centering
    \includegraphics[width=.32\textwidth]{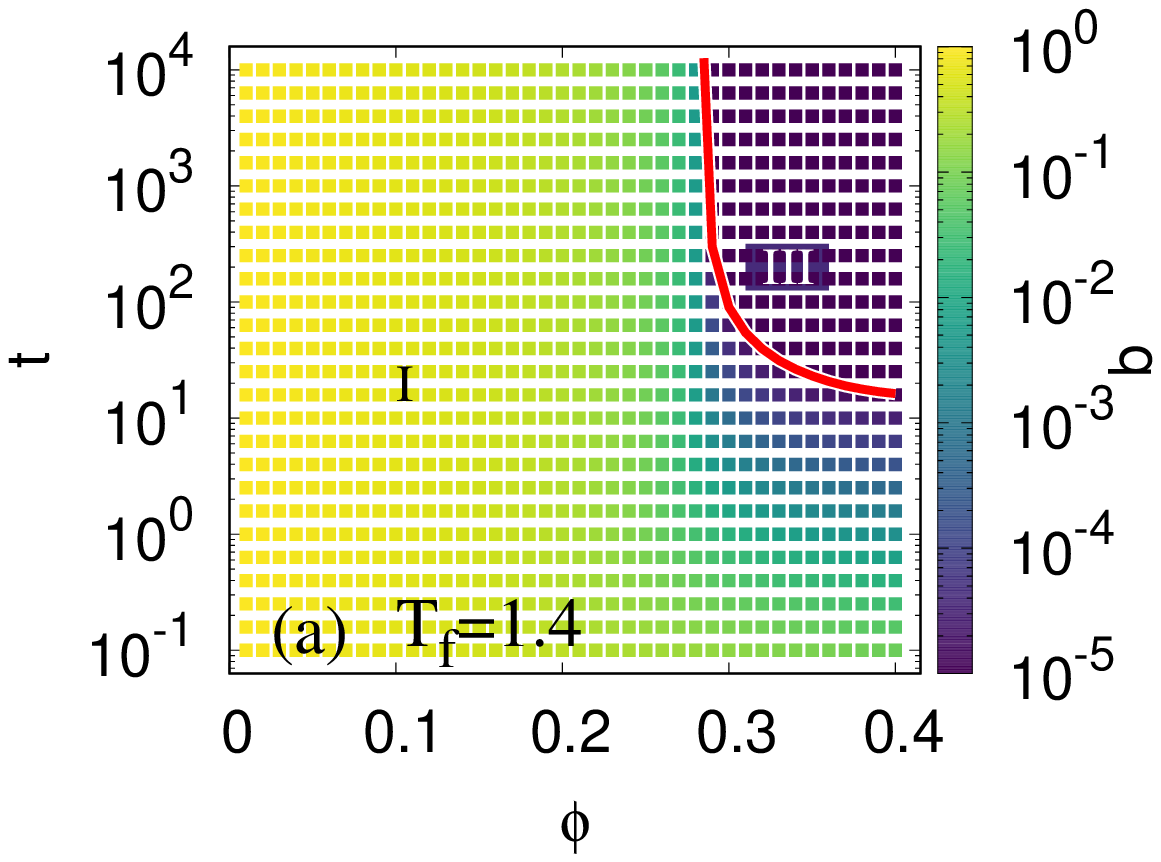}
    \includegraphics[width=.32\textwidth]{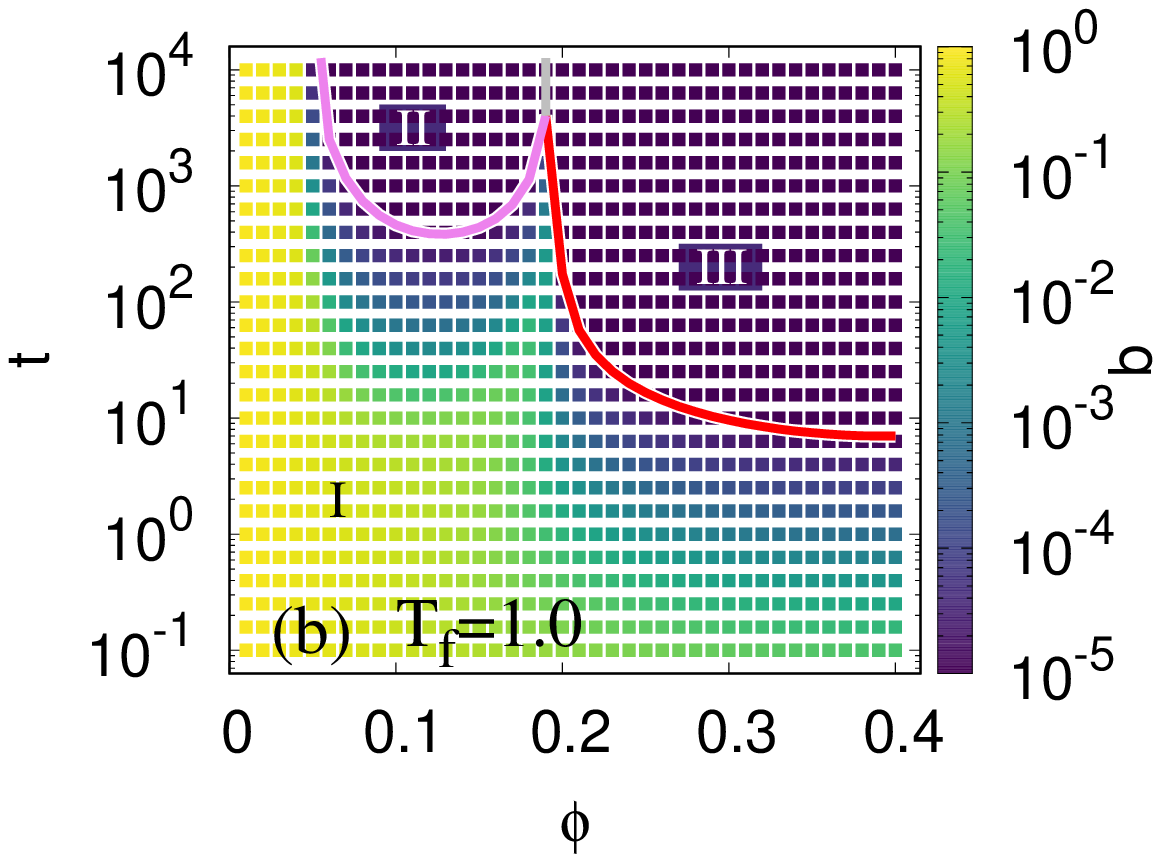}
    \includegraphics[width=.32\textwidth]{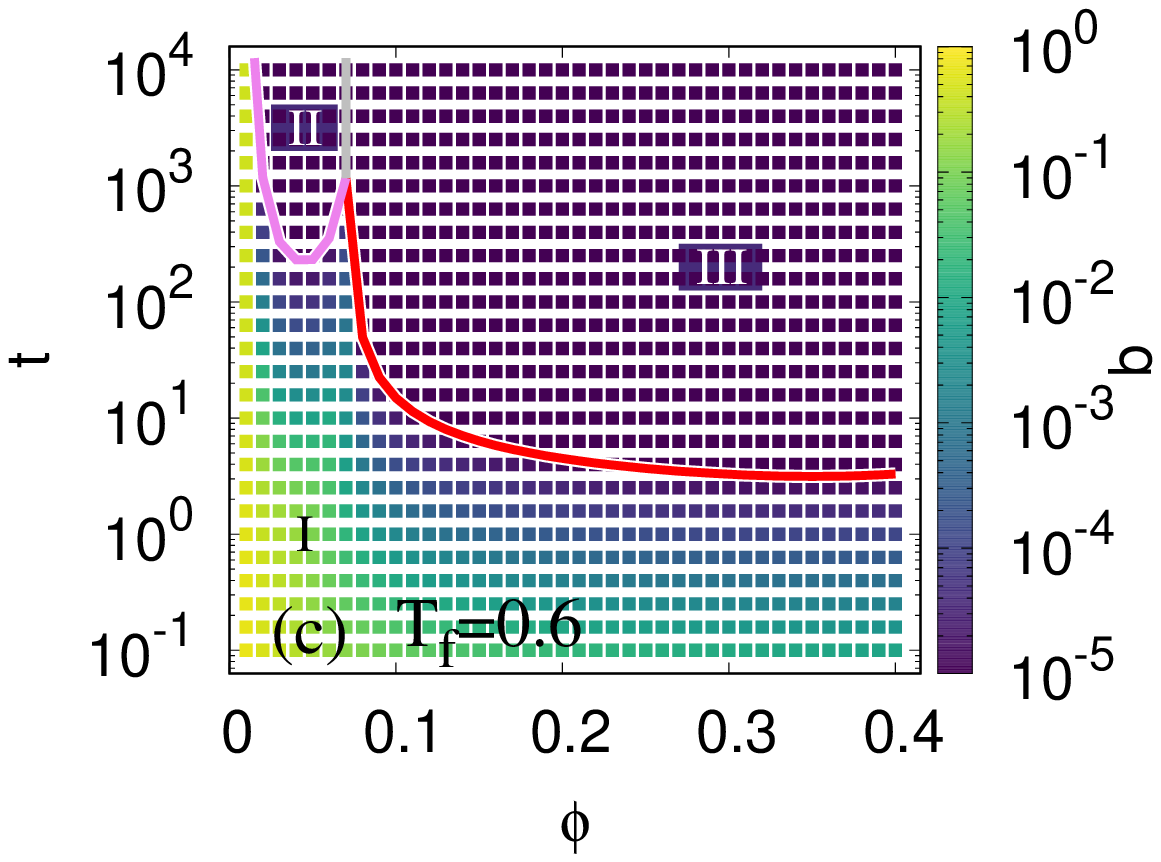}    
    \includegraphics[width=.32\textwidth]{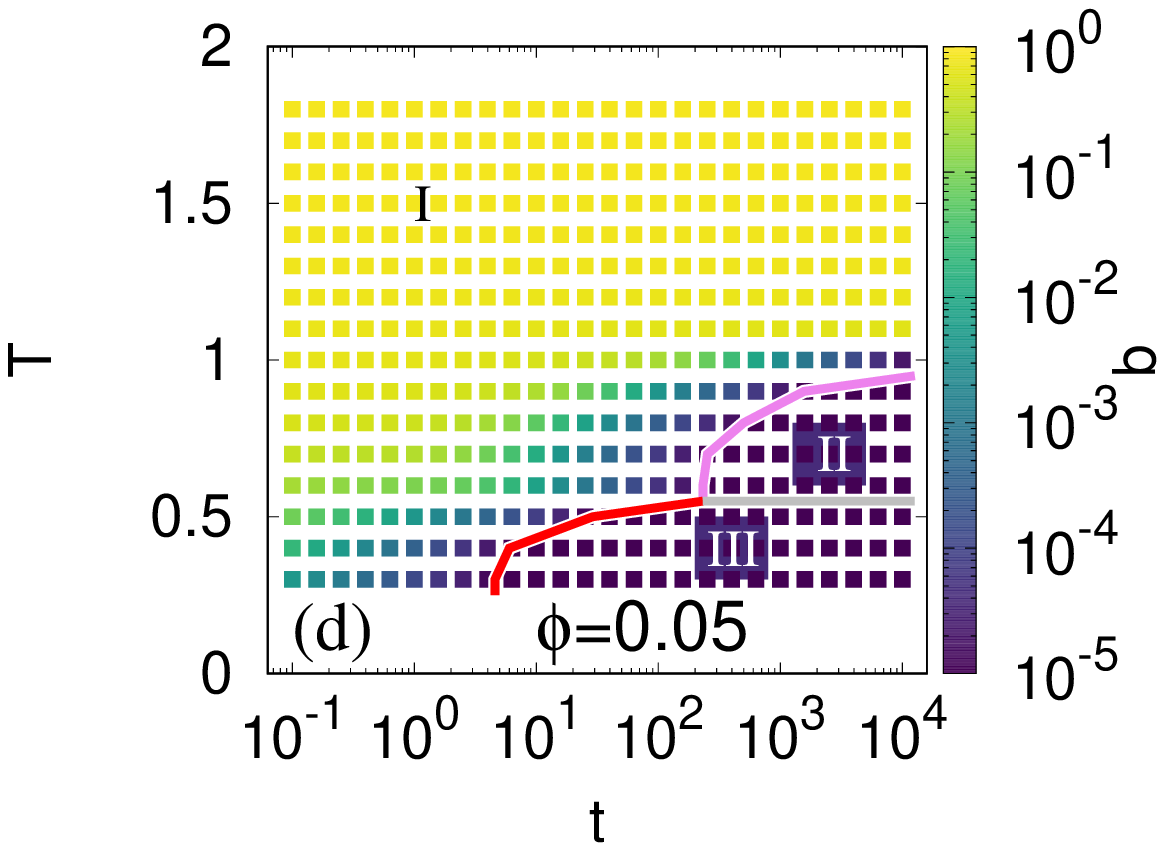}
    \includegraphics[width=.32\textwidth]{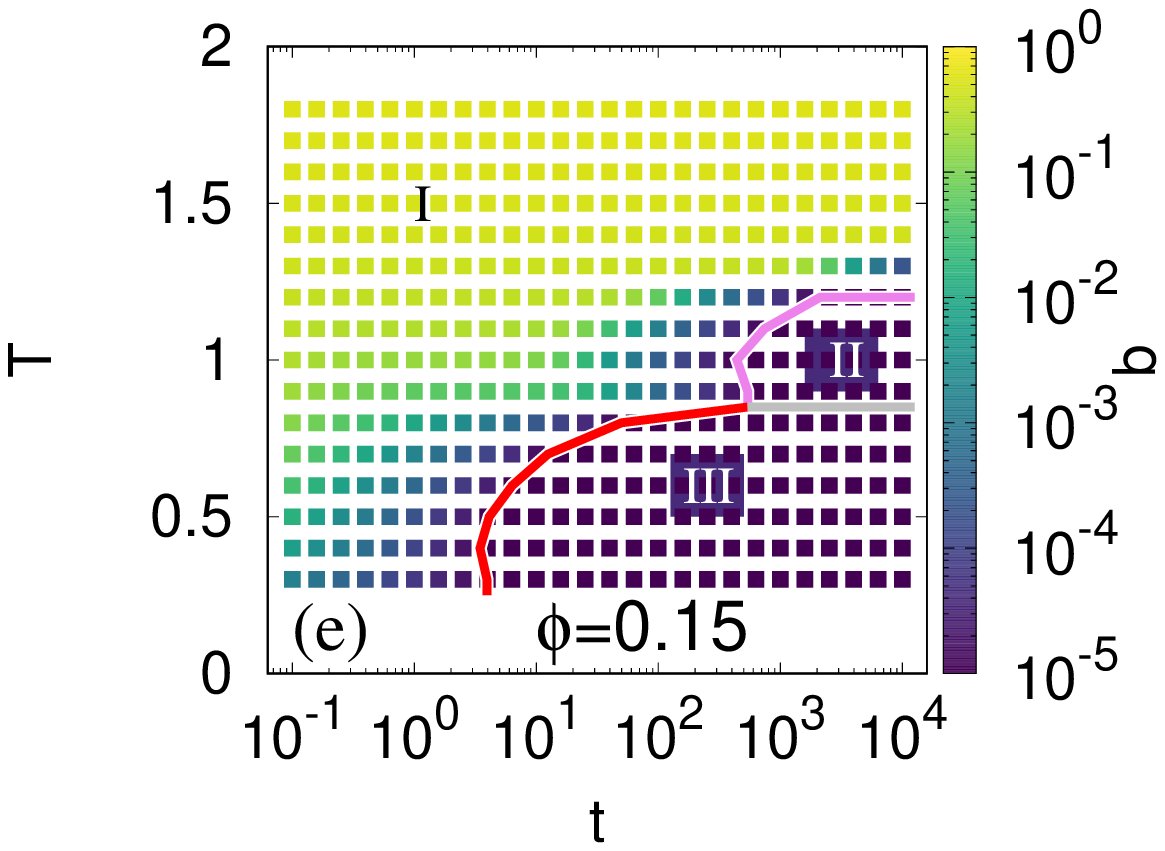}
    \includegraphics[width=.32\textwidth]{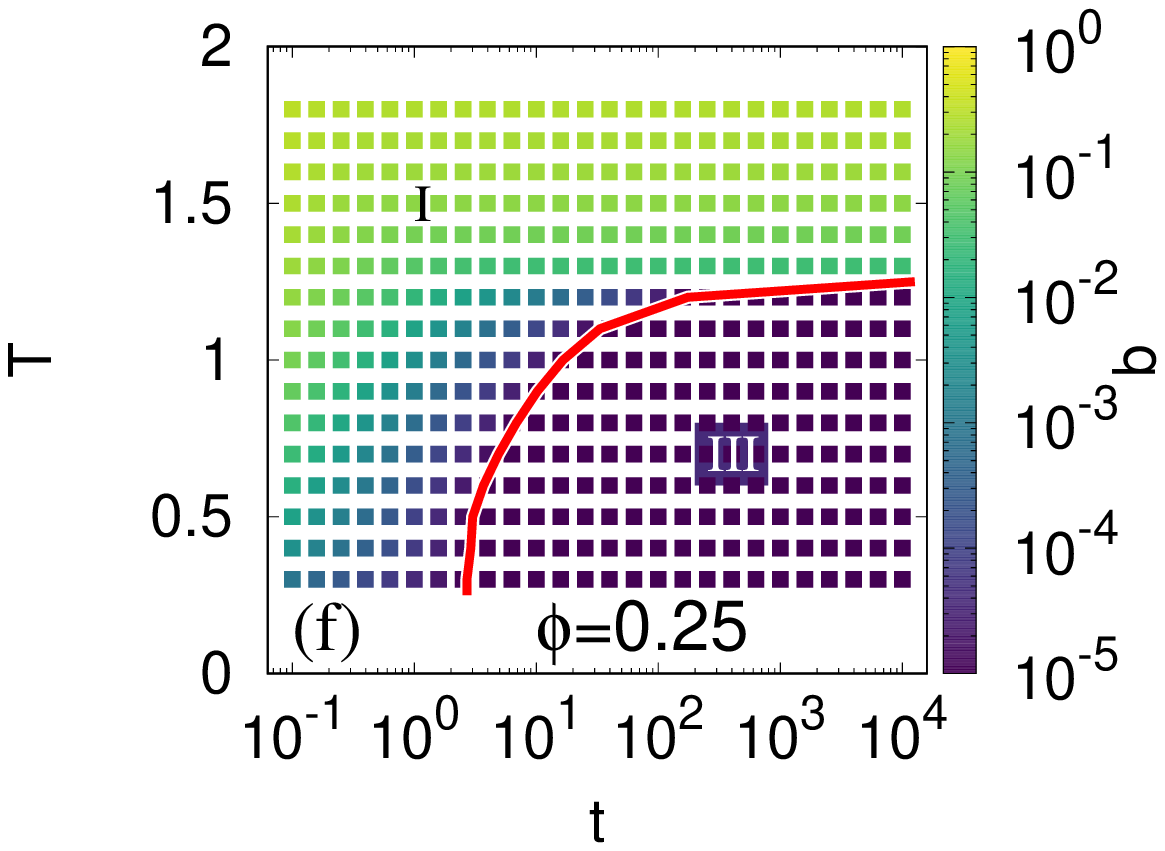}    
   \caption{$\phi$-$t$ planes at constant $T_f=1.4$ (a), $1.0$ (b) and $0.6$ planes and $t$-$T$ planes at constant $\phi=0.05$ (d), $\phi=0.15$ (e) and $\phi=0.25$ (f)  of $b(t;\phi,T)$. Fluid states are expected at region I in each of the diagrams, with $b(t;\phi,T)>10^{-5}$ boundary condition. The solid red and pink lines corresponds to the $b(t;\phi,T)=10^{-5}$ boundary, while segregating glass-like (red enclosing region III) from the gel-like (pink enclosing region II) arrested states.}
\label{fig4}
\end{figure*}

A similar but noticeably slower evolution is exhibited by the iso-$b$ line corresponding to $b_0=  10^{-5}$ (solid line). As  illustrated by Figs.  \ref{fig3}(d)-(f), it is only until  $t=10$ that this line has moved well into the $(\phi,T)$-window studied, and its difference with the equilibrium limit $b(t\to\infty;\phi,T)=b^{eq}(\phi,T) = 10^{-5}$ (thick solid bright gray line) is still observable at  $t=10^{3}$. Fig.  \ref{fig3} thus describes the continuous  time-evolving pattern of mobility distribution on the $(\phi,T)$ state space, in which the time-dependent iso-$b$ lines serve as sharp but artificial visual aids.  At even  longer times, however, this evolution  becomes less and less perceptible, to the point that it eventually appears  stationary. In fact, this already occurs at the waiting time $t=10^5$, whose snapshot coincides, within the resolution of the figure, with the non-equilibrium glass transition diagram $b(t\to\infty;\phi,T)=b^{eq}(\phi,T)$ of Fig.  \ref{fig1}(c). Similarly, the iso-$b$ line $b(t=10^5;\phi,T) = 10^{-5}$ is  indistinguishable from its long-time limit $b^{eq}(\phi,T) = 10^{-5}$, and both coincide in practice with the solid line, which is the dynamic arrest line $b^{eq}(\phi,T) = 0$.

The slow dynamic features illustrated in Figs.  \ref{fig3}(d)-(f) and \ref{fig1}(c) help us understand the nature of the experimental glass transition. A conventional  empirical criterion \cite{angellreview1,hunterweeks} defines a glass as a supercooled liquid whose  $\alpha$-relaxation time  $\tau_\alpha$ (or its viscosity $\eta$) exceeds a large arbitrary threshold $\tau^{(g)}_\alpha$ (or $\eta^{(g)}$). Since $\tau_\alpha(t;\phi,T) \propto b(t;\phi,T)^{-1}$   \cite{nescgle3}, we propose our ``empirical'' definition of an  arrested  state as one whose mobility $b$ has dropped not to 0, but below a very small but finite threshold $b^{(g)}$. If we now take $b^{(g)}$ as $10^{-5}$  (having in mind colloidal glasses \cite{hunterweeks}), we then conclude that the evolution of the solid line in Figs.  \ref{fig3}(d)-(f) and \ref{fig1}(c) represents the non-equilibrium evolution of such an ``empirical'' glass transition line. 

The whole evolution in Figs.  \ref{fig3}(d)-(f) and \ref{fig1}(c), with its beginning, development, and end, is what we refer to as a \emph{time-dependent non-equilibrium phase diagram}. Its kinetic perspective, and its description of dynamically-arrested phases, constitute the most relevant difference with respect to ordinary equilibrium phase diagrams, whose proper counterpart is the non-equilibrium glass transition diagam in Fig. \ref{fig1}(c), now obtained as the $t\to\infty$ limit of this $t$-dependent process.

One important question refers to the possible dependence of the kinetic scenario just presented, on the value of $b^{(g)}$ when this arbitrary parameter is decreased even further. We found that the basic scenario is essentially the same, except that in order to observe qualitative features similar to those displayed in Figs.  \ref{fig3}(d)-(f) and \ref{fig1}(c), much longer waiting times and much higher resolution in $b(t)$, $\phi$ and $T$, will be required. What will not change, however, are the long-time limiting lines $b(t\to\infty;\phi,T)=b^{eq}(\phi,T) = b^{(g)}$ for $ b^{(g)}$ smaller than $10^{-5}$, all of which virtually superimpose (again, within the resolution of the figure) on the solid line of Fig. \ref{fig1}(c). Thus, this line is the stationary dynamic arrest line $b^{eq}(\phi,T) = 0$, one of the most relevant elements of the non-equilibrium glass transition diagram, first presented and explained in detail in Ref.  \cite{nescgle5}.

\subsection{Isochoric and isothermal cuts of $b(t;\phi,T)$.}\label{subsect3.3}

Let us now display the same information contained in $b(t;\phi,T)$, but this time along the plane $(t,\phi)$ at fixed final temperature $T$, as in Fig. \ref{fig2}(b), and along the plane $(T,t)$ at constant $\phi$, as in Fig. \ref{fig2}(c). The former is done in Figs. \ref{fig4}(a)-(c) for a supercritical ($T=1.4$) and two subcritical ($T=1.0$, $0.6$) final temperatures, and the latter in Figs. \ref{fig4}(d)-(f) for the isochores $\phi=0.05$, $0.15$ and $0.25$. In each of these figures, the intersection of the respective plane with the empirical dynamic arrest transition surface  $b(t;\phi,T)= b^{(g)}=10^{-5}$ is represented by the colored solid lines, which indicates the time it takes for the state of the system to pass from fluid  ($b(t;\phi,T)>b^{(g)}$) to arrested ($b(t;\phi,T)< b^{(g)}$), and where the red solid line indicates a transition to the expected glass-like states (region III) and the pink solid line to the expected gel-like states (region II).

From the isothermal diagrams in Figs. \ref{fig4}(a)-(c), it can be seen that, at a given isotherm, we only have two or three possible final states, depending on the temperature. Thus, at  supercritical temperatures (Fig. \ref{fig4}(a)) the system is expected to reach equilibrium at low volume fractions, but above a critical volume fraction $\phi_g(T)$ that depends on temperature ($\phi_g(T=1.4)\approx 0.29$), a transition will occur to an arrested state corresponding to a high-density hard-sphere like (i.e., ``repulsive'') glass. In fact, the high-$T$ limit of $\phi_g(T)$ is precisely $\phi_g(T=\infty) \approx 0.58$.

In contrast, for subcritical temperatures, illustrated by Figs. \ref{fig4} (b) and (c), a third possibility emerges, namely, the formation of arrested sinodal-decomposition gel states, corresponding to region II in the non-equilibrium glass transition diagram of   Fig. \ref{fig1}(c). In fact, for  isotherms only slightly subcritical (not illustrated here), the gel transition will be confined to a small volume fraction interval around the volume fraction $\phi_c$ of the critical point, and will occur only after a very long waiting time, so that it would not appear in the time-window of these figures. At the lower temperatures illustrated in these figures, however, this transition grows wider in its volume fraction interval, and occurs earlier in time, so that it now appears in our illustrative time-window. Simultaneously, the volume fraction  $\phi_g(T)$ above which the fluid becomes a hard-sphere glass, decreases, leading to the long-time merging of the volume fraction intervals corresponding to gel and to glass formation. The results also exhibit the fact that  the transition to a glass occurs earlier than the formation of gels.

\begin{figure*}
\begin{center}
\includegraphics[width=.35\textwidth]{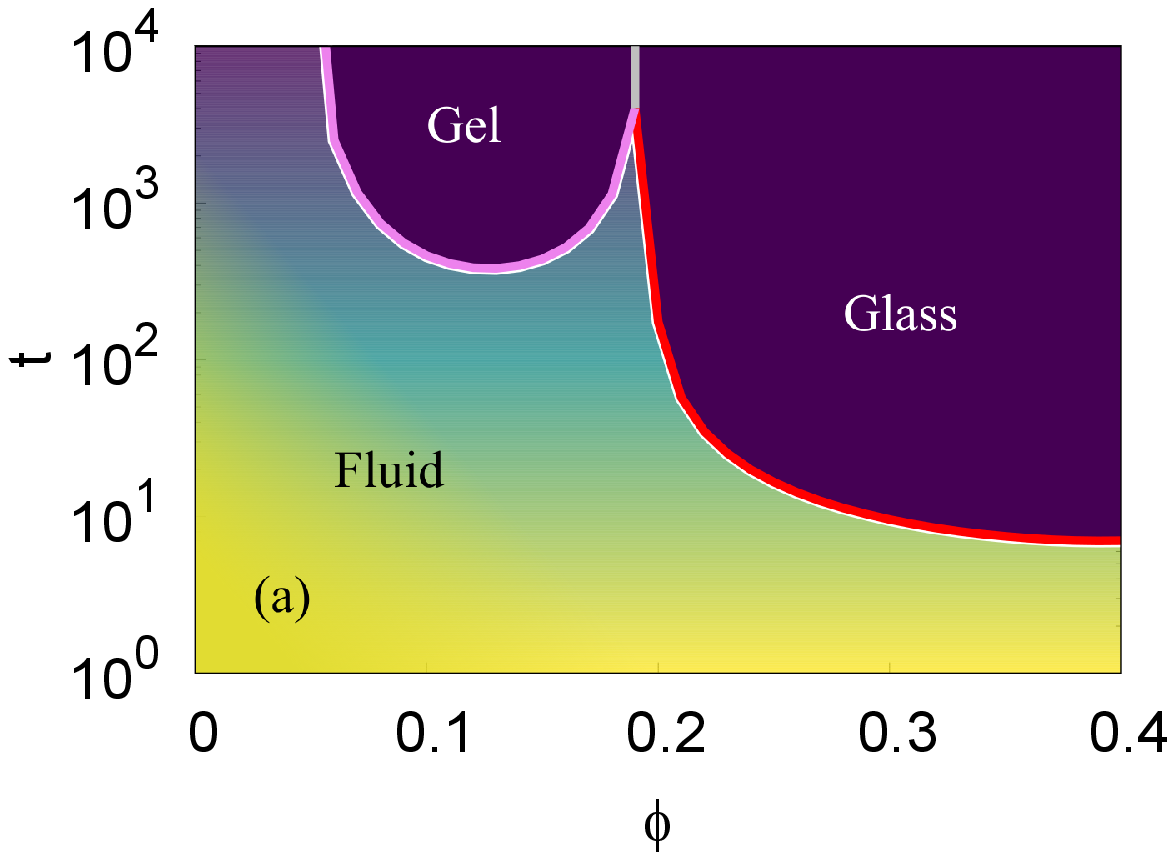}
\includegraphics[width=.35\textwidth]{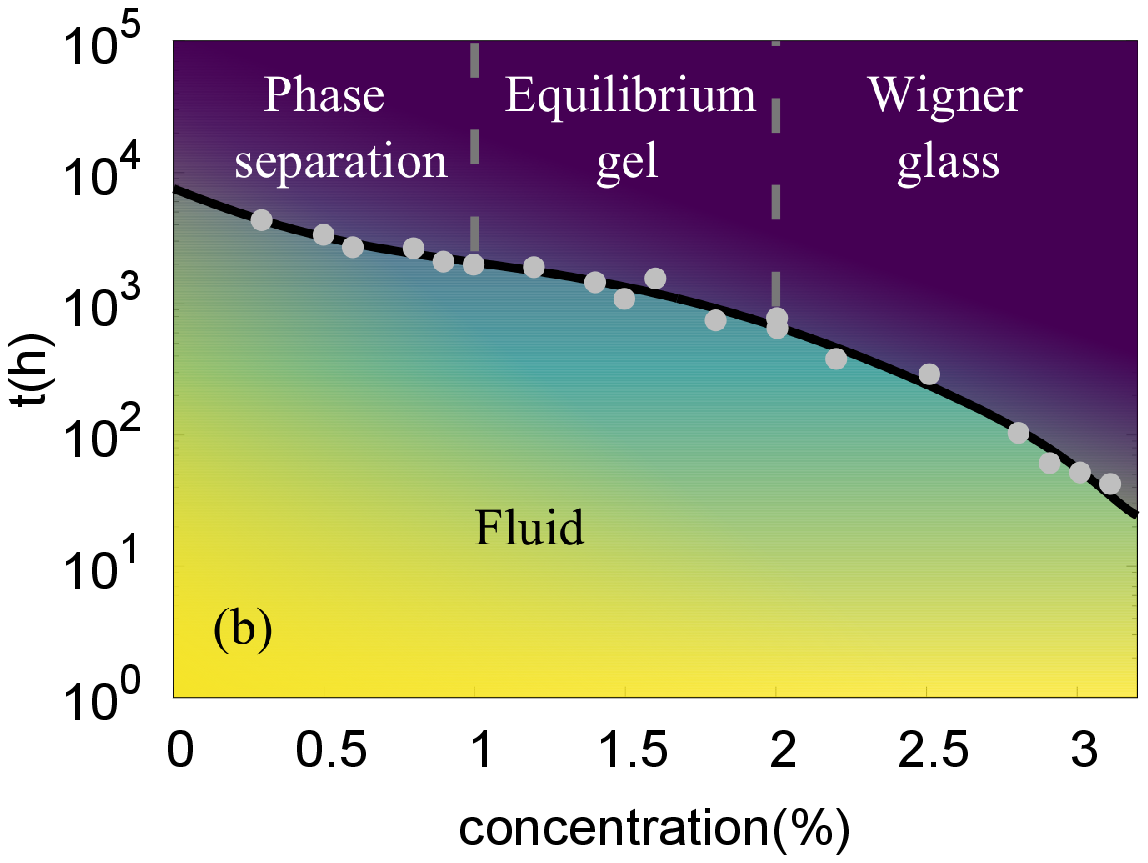}
\includegraphics[width=.35\textwidth]{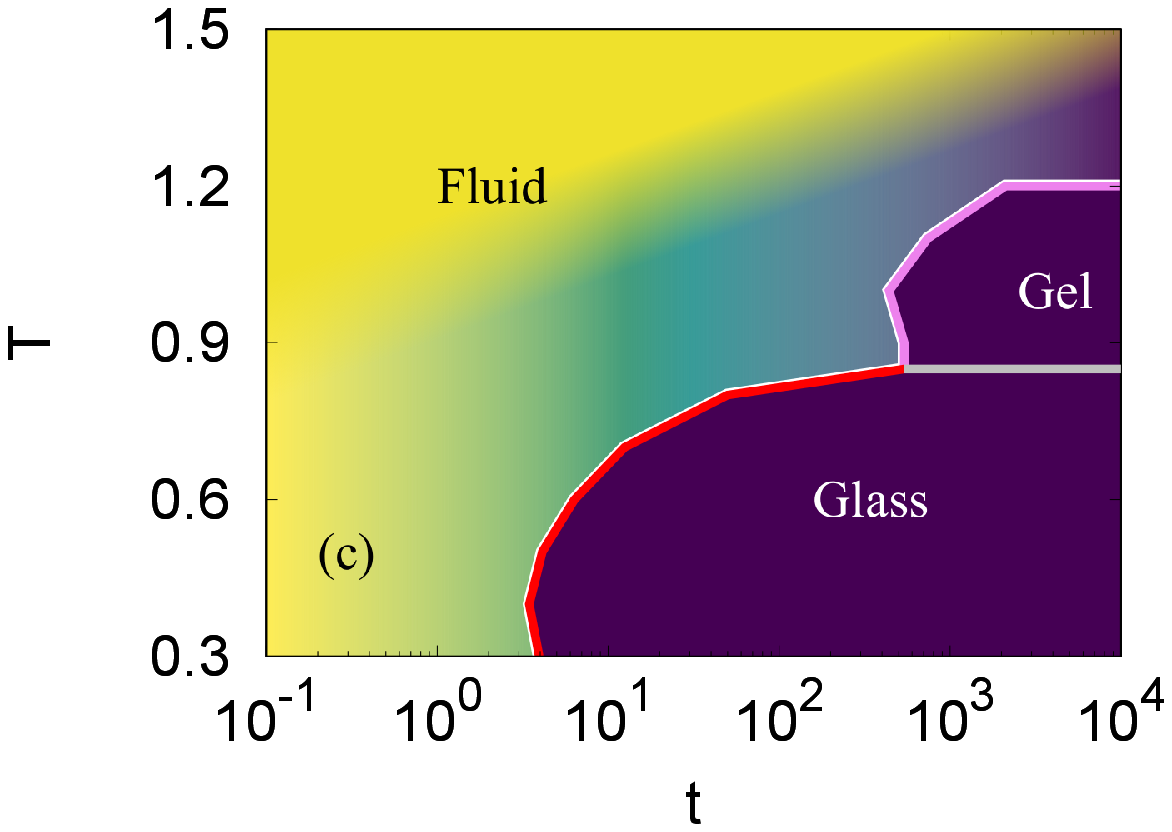}    
 \includegraphics[width=.35\textwidth]{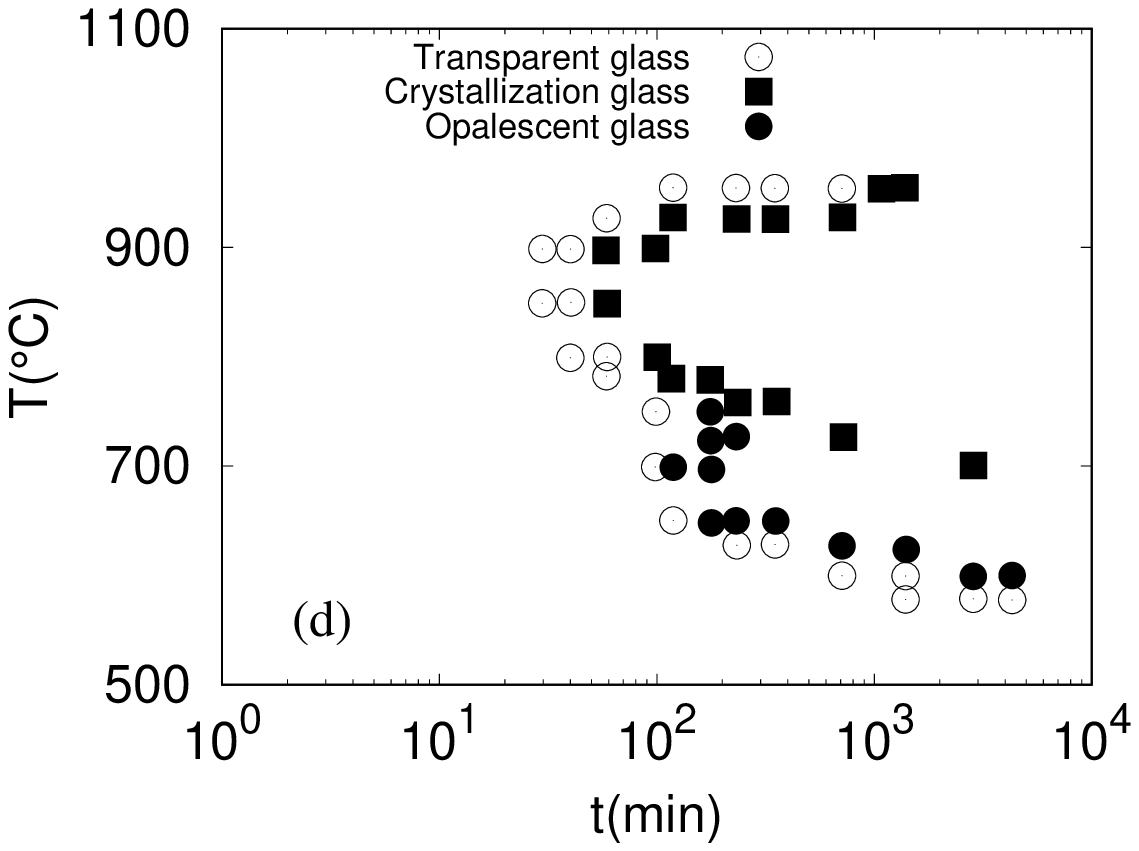}    
\caption{(a) Time-dependent non-equilibrium phase diagram for the HSSW system at $T=1.0$. (b) Laponite time dependent diagram sketched from the experimental data in Ref. \cite{ruzicka1} where time is measured in hours after filtering the sample which was inmediatly obtained after 30 minutes of stirring, and concentration is weight concentration. (c) Time-dependent non-equilibrium phase diagram for the HSSW system at $\phi=0.15$. (d) Non-equilibrium state transformation of the borosilicate glass $65$SiO$_2\cdot25$B$_2$O$_3\cdot10$NA$_2$O, reproduced from the time-temperature transformation (TTT) diagram in fig. 1 of Ref. \cite{nakashima}}
\label{fig5}
\end{center}
\end{figure*}

Similar considerations can be made regarding the evolution of $b(t;\phi,T)$ along planes of constant volume fraction, presented in Figs. \ref{fig4} (d)-(f)  ($\phi=0.05,\ 0.15$, and 0.25). The results for these three isochores illustrate the fact that  the deepest quenches always lead to glass formation. For deepest quenches we mean $T$ below the temperature $T_c(\phi)$ of the gel-glass transition (the dashed line of Fig. \ref{fig1}(c)). However, shallower quenches may lead to the formation of gels  (for $T$ above $T_c(\phi)$ but below the spinodal temperature $T_s(\phi)$) or to the system reaching its corresponding equilibrium state  (for $T$ above $T_s(\phi)$). Gel formation, however, does not occur for isochores with volume fraction above the bifurcation point (at which $T_c(\phi)=T_s(\phi)$), as illustrated by Fig. \ref{fig4}(f).

\section{Summary and perspectives.}\label{sect4}

In the previous section we have  illustrated the applicability of the NE-SCGLE theory  to predict from first principles what we refer to as the waiting-time dependent non-equilibrium phase diagram of a simple model liquid. This is represented in essence by the non-equilibrium evolution of  the mobility function $b(t;\phi,T)$, the simplest but most meaningful dynamic order parameter emerging from the solution of the NE-SCGLE equations (\ref{relsigmadif2pp}) and (\ref{bdt})-(\ref{lambdadk}). 

The analysis just presented complements recent work, in which the predicted behavior of $b(t;\phi,T)$ during \emph{individual} quench experiments, revealed intriguing latency effects \cite{nescgle8}, previously observed experimentally \cite{guoleheny}, as well as other experimentally-observed structural signatures of gel formation \cite{nescgle7}. In contrast with this individual quench experiments, the concept of non-equilibrium phase diagram is based on an ensemble of instantaneous quenches. This concept is meant to describe theoretically, experiments in which one prepares an ensemble of samples that differ in the final density and temperature of the quench, and then simultaneously monitor the irreversible evolution of each sample. In contrast with equilibrium phases, whose definition relies solely on thermodynamic and structural order parameters, non-equilibrium arrested phases must be described in terms of  time-dependent \emph{dynamic} order parameters, such as $b(t)$. 

As illustrated in Fig. \ref{fig3}, the scenario that emerges from the evolution of $b(t;\phi,T)$ describes the  transient state of the ensemble, which reveals the obvious, but sometimes subtle, kinetic differences between different regions of state space. In particular, it reveals the apparent difference in the time it takes the system to form a glass or a gel. To dwell into its analysis, the same information was then presented in Fig. \ref{fig3} for sub-ensembles at various volume fractions but same final temperature or at a given  volume fraction but various final temperatures. As it happens, however, these two formats actually correspond to real practical experimental protocols to describe the  waiting time dependence of the state of a liquid or melt after preparation or after a temperature quench.

To illustrate its experimental pertinence, in Fig. \ref{fig5} we compare these two formats of the NE-SCGLE theoretical $t$-dependent phase diagram, side by side with two examples of experimental diagrams that report the time dependence of the formation of non-equilibrium arrested phases in precisely those two formats. It should be clear, however, that this comparison does not refer to the detailed non-equilibrium phases formed, since neither of these experimental systems is actually represented by the HSSW model). In the first example, shown in Fig. \ref{fig5} (b), we schematically reproduce the time-concentration diagram reported in Ref. \cite{ruzicka1} to describe the aging of an ensemble of  Laponite suspensions prepared at various concentrations and the same temperature. The solid line indicates the time needed by the system to phase separate or become dynamically arrested into a gel or a glass phase, depending on concentration.

%This is illustrated in Fig. \ref{fig5} with two examples of experimental diagrams that exhibit the time dependence of the formation of non-equilibrium arrested phases. In the first example, shown in Fig. \ref{fig5} (a), we schematically reproduce the time-concentration diagram reported in Ref. \cite{ruzicka1} to describe the aging of an ensemble of  Laponite suspensions prepared at various concentrations and the same temperature. The solid line indicates the time needed by the system to phase separate or become dynamically arrested into a gel or a glass phase, depending on concentration.

Clearly, the format of this report is the same as that of the $t-\phi$ diagrams in Figs. \ref{fig4}(a)-(c), although with quite apparent or rather subtle differences. For example, as explained in Ref. \cite{ruzicka2}, the experimentally measured dynamic order parameter was  the $\alpha$-relaxation time, and not directly the mobility function. In our HSSW illustrative example, both order parameters are provided by the solution of the NE-SCGLE equations, and both have essentially the same behavior. The main differences between this precise experimental diagram and the theoretical  diagrams in Figs. \ref{fig5}(a)-(c) is, however, the fact that our simple HSSW model system was not meant to represent the relevant interactions between Laponite particles, which are electrically charged and have a coin-shaped hard core. A direct application of the non-equilibrium theoretical methodology developed here to a more realistic interaction potential is, in principle, perfectly possible, but it constitutes a separate project. 

The second example, shown in Fig. \ref{fig5} (d), corresponds to the expected time-dependent non-equilibrium phases of a borosilicate glass, reproduced from a time-temperature-transformation (TTT) diagram reported by Nakashima et al. in Ref. \cite{nakashima}. Here the format of the report is analogous to the format of the time-temperature diagrams in Figs. \ref{fig4}(d)-(f), and in both cases, the interference between spinodal decomposition and dynamical arrest is a relevant feature. The differences, however, are stronger and more fundamental. These include the molecular nature of the multicomponent experimental system, vs. the Brownian nature of our illustrative monocomponent model, and the fact that the spinodal conditions correspond in the experiment to a demixing thermodynamic instability, and not to a gas-liquid phase separation instability. And, of course, that the experiment reflects the interference with crystallization, not considered yet in the theory, and that the interparticle interactions of the HSSW model do not represent the strong directional bindings between atoms in the experimental system.

Nevertheless, there is no fundamental impediment for the eventual incorporation of each of these effects in extended versions of the NE-SCGLE theory and by changing at will the details of the microscopic interaction potential $u(r)$. The limiting step is, thus, no longer the absence of a fundamental physical framework to address these challenges, but the need to adapt the theory to the detailed microscopic modeling of each specific experimental condition of interest. For example, still within the HSSW model, from the perspective of the colloid and soft condensed matter community, much shorter-ranged attractions are of greater interest than the regime $(\lambda-1)= 0.5$ studied here. A thorough and  systematic study of shorter-ranged attractions is the subject of current research that we hope to report separately. In fact, in the appendix we provide a brief summary of the main changes of the non-equilibrium glass-transition diagram, as the regime  $(\lambda-1)\ll 0.5$ is approached. Beyond the HSSW model, however, we believe that the advances already made in extending the  NE-SCGLE theory to multicomponent systems  \cite{nescgle4} and to systems of non-spherical particles \cite{gory1,nescgle7}, open the route to the description of more subtle and complex non-equilibrium amorphous states of matter.

At this moment such advances would only involve the present version of the NE-SCGLE theory  (Eq. (\ref{relsigmadif2pp}), together  with Eqs. (\ref{bdt})-(\ref{lambdadk})). This theory, however, is still susceptible of deeper revision regarding its fundamental basis and current limitations. For example, as explained in detail in Ref. \cite{nescgle1} (and summarized in Sect. II of Ref. \cite{nescgle6}), the present NE-SCGLE  equations  result from neglecting the spatial heterogeneities represented by the deviations $\Delta \overline{n}(\textbf{r},t)$ of the mean value $\overline{n}(\textbf{r},t)$ of the instantaneous local density $n(\textbf{r},t)$ from its uniform bulk value $n$. This approximation,   imposed on the original spatially non-uniform equations derived in Ref. \cite{nescgle1}, leads to Eq. (\ref{relsigmadif2pp}) and Eqs. (\ref{bdt})-(\ref{lambdadk}) above. Restoring the time- and space-dependence of $\overline{n}(\textbf{r},t)$ involves mostly a technical obstacle, but its implementation will provide the direct visualization of the morphological evolution of the glass-forming liquid as it becomes a non-equilibrium amorphous solid. At least to linear order in the deviations $\Delta \overline{n}(\textbf{r},t)$ , this is perfectly possible, as we shall report separately \cite{beni2}. 

Similarly, it would be highly desirable to go beyond the local stationarity approximation, upon which the NE-SCGLE theory is based. This approximation models the spontaneous fluctuations $\delta \overline{n}(\textbf{r},t+\tau)$ as a momentarily stationary process during a correlation time $\tau$, but as a globally non-stationary process regarding its dependence on the waiting time $t$. In contrast, Latz's non-equilibrium MCT theory does not appeal to this approximation, thus being are more general, but also less practical, than the current NE-SCGLE theory. In fact, we are not aware of concrete applications of Latz's proposal to specific  examples of structural glasses. Thus, while it will always be desirable to pursue Latz's more ambitious program, another interesting possibility is to actually introduce  the local stationarity approximation in the non-equilibrium MCT equations, most likely leading to predictions similar to  the list of successful applications of the NE-SCGLE theory. This, however, is at this moment only another relevant perspective of fundamental research aimed to achieve the most accurate understanding of non-equilibrium amorphous solids.

\section*{Author's Contributions}
All authors contributed equally to this work.

\begin{acknowledgments}
We gratefully acknowledge helpful discussions with Drs. Pedro Ram\'{\i}rez-Gonz\'alez, Leticia L\'opez-Flores, Ernst van Nierop, Ricardo Peredo-Ortiz. We specially acknowledge the advice of Dr. Luis F. Elizondo-Aguilera and Prof. Thomas Voigtmann, and of two anonymous referees. We are also grateful to Consejo Nacional de Ciencia y Tecnolog\'{\i}a (CONACYT, M\'exico) for financial support through grants No. CB A1-S-22362 and LANIMFE 314881, and a graduate fellowship to J.B. Z.-L. 
\end{acknowledgments}

\section*{Data availability}

The data that supports the findings of this study can be reproduced through the numerical methods and from the public repository: \url{https://github.com/LANIMFE/HS\_HSSW\_quench\_NESCGLE}, and may also be available from the corresponding author upon request. Details upon the employed numerical methods are discussed in appendix C.

\nocite{*}

\appendix

\section{Equilibrium properties.}\label{appendix1}

To better appreciate the relationship between the conventional equilibrium statistical thermodynamic theory of fluids \cite{mcquarrie, hansen} and its non-equilibrium extension provided by the NE-SCGLE theory of irreversible processes \cite{nescgle1}, this Appendix summarizes  some elements of integral equation theory \cite{hansen} and the equilibrium density functional theory (DFT) of liquids \cite{evans}, and introduces one particular approximation (referred to as modified mean field approximation). Although this Appendix basically reviews well-established material \cite{mcquarrie, hansen}, that specialists of the theory of liquids might simply skip, here we present it in the format that best suits the purpose of its extension to non-equilibrium conditions.

\subsection{The modified mean field (MMF) approximation.}\label{appendix1.1}

Within a perturbative spirit illustrated, for example, by Foffi et al. \cite{foffidawson2002} for model systems whose pair potential is the sum of a hard-sphere  repulsion term $u_{HS}(r)$ plus an attractive tail $u_A(r)$, let us approximate the free energy $F=\mathcal{F}[n;T]$ of our system by the superposition 
\begin{equation}
\mathcal{F} [n,T]=\mathcal{F}_{HS} [n,T] + \mathcal{F}_{A} [n,T],
\label{pertF}
\end{equation} 
of its \emph{exact} HS value $\mathcal{F}_{HS} [n,T]$, plus a  contribution $\mathcal{F}_{A} [n,T]$   of the attractive interactions. In this work the latter will be approximated by its \emph{modified mean field} (MMF) approximation  \cite{groh,teixeira,frodldietrich,tavares}, defined by the following functional form, 
\begin{equation}
\mathcal{F}_{A} [n,T]= - \frac{k_BT}{2}\int d \mathbf{r} \ \int d \mathbf{r}' n(\mathbf{r}) 
f_A(\mid \mathbf{r}-\mathbf{r}'\mid;T) n(\mathbf{r}'),
\label{Fammf}
\end{equation}
where 
\begin{equation}
f_A(r;T) \equiv \left\{ \exp [-\beta u_A(r)] -1\right\} \theta (r-\sigma).
\label{fammf}
\end{equation}

This implies that we can approximate the second functional derivative $\mathcal{E}[\mid \mathbf{r}-\mathbf{r}'\mid; n,T] \equiv  \left( \delta^2 \mathcal{F}[n,T]/k_BT/\delta n(\mathbf{\mathbf{r}})\delta n(\mathbf{\mathbf{r}'}) \right)$ by the superposition  
\begin{equation}
\mathcal{E}(r;n,T)=\mathcal{E}_{HS}(r/\sigma;\phi)+ f_A(r;T),  
\label{superposition}
\end{equation} 
where $\mathcal{E}_{HS}(r/\sigma;\phi)$ is the exact hard-sphere value of $\mathcal{E}(r;n,T)$, also written as 
\begin{equation}
\mathcal{E}_{HS}(r/\sigma;\phi)=\delta (\mathbf{r})/n-c_{HS}(r/\sigma;\phi), 
\label{chsrphi}
\end{equation}  
which implies that we can approximate  $\mathcal{E}(r;n,T)$ by  
\begin{equation}
\begin{split}
\mathcal{E}(r;n,T)=\delta (\mathbf{r})/n-c_{HS}(r/\sigma;\phi) - \left\{ \exp [-\beta u_A(r)] -1\right\} \times \\
 \theta (r-\sigma),
\end{split}  
\label{superposition2}
\end{equation} 
where $c_{HS}(r/\sigma;\phi)$ is the exact direct correlation function of the HS fluid. 

In practice, for $c_{HS}(r/\sigma;\phi)$ we will adopt the analytical and virtually exact expression $c_{PY/VW}(r/\sigma;\phi)$ provided by the Percus-Yevick approximation with the Verlet-Weis correction \cite{percusyevick,verletweiss}. With this approximate determination of the thermodynamic input we can proceed to the solution of the NE-SCGLE equations, whose results are presented and discussed in Sect. \ref{sect3}.

\subsection{Equilibrium phase diagram.}\label{appendix1.2}

According to the conventional theory of liquids \cite{hansen,evans}, all the equilibrium thermodynamic information, including the equilibrium phase diagram, can be derived from the Helmholtz free energy density-functional $F=\mathcal{F} [n,T]$. For example, from the compressibility equation \cite{hansen}, 
\begin{equation} 
\left(\frac{\partial  [\beta p (n,T)]}{\partial n}\right)_T= n \mathcal{E}(k=0;n,T), 
\label{compreq}
\end{equation}
we can compute the mechanical equation of state of our HSSW model. Denoting the dimensionless pressure $[\sigma^3 p /\epsilon]$ simply by $p$, such an equation reads
\begin{equation}
\begin{split}
p(\phi,T)= p_{HS}(\phi,T) - \frac{24}{\pi}\left( \lambda^3  -1 \right)\phi^2 T\times \\ 
\left( e^{1/T} - 1 \right) ,
\end{split}
\label{PHSSW}
\end{equation}
with $p_{HS}(\phi,T)$ being the pressure of the reference HS system, given by  the Carnahan-Starling equation of state, 
\begin{equation} 
p_{HS}(\phi,T)\equiv \frac{6 T\phi ( 1 + \phi + \phi^2 - \phi^3 ) }{\pi \left( 1 - \phi \right)^3}.
\label{PHS}
\end{equation}

Using Eqs. (\ref{PHSSW}) and (\ref{PHS}) plus the condition $(\partial p/\partial n)_T=0$ one can determine the spinodal line of the gas-liquid transition, while the binodal line can be obtained through Maxwell  construction \cite{callen}. The corresponding results of this exercise for the HSSW liquid with $\lambda=1.5\sigma$ are illustrated in Fig. \ref{fig1}(a), with the spinodal, binodal and freezing lines having the conventional thermodynamic meaning \cite{callen}. In Fig. \ref{fig1}(a), however, we have also plotted the gray solid line along which the height $S^{eq}(k_{max};n,T)$  of the main peak of the equilibrium static structure factor $S^{eq}(k;n,T)= 1/n\mathcal{E}(k;n,T)$ remains constant and equal approximately to 3.1. This line, according to the phenomenological rule of crystallization of Hansen and Verlet \cite{hansenverlet}, coincides approximately with the freezing transition. Next to it we also draw the closest iso-diffusivity gray dashed line, along which the long-time self-diffusion coefficient $D_L(n,T)$, normalized by its short-time value $D_0$, is constant, $b^{eq}(n,T)\equiv D_L(n,T)/D_0 \approx 0.1$. According to L\"owen's phenomenological  dynamic criterion of freezing \cite{lowen}, this line should also lie close to the freezing line.

This allows us to have a semiquantitative sketch of important elements of the equilibrium phase diagram, such as the location of the boundary of stability of equilibrium fluid states, which lie to the left and above this iso-diffusivity line and above the binodal curve. To the right and below the freezing line we should have the non-uniform equilibrium coexistence of the liquid with the crystalline solid or, if crystallization is suppressed, the supercooled metastable equilibrium liquid. In  the region between the binodal and the spinodal lines we either have non-uniform gas-liquid equilibrium coexistence or metastable uniform fluid states.

\section{SCGLE and NE-SCGLE expressions, asymptotic limits and glass transition diagrams for the HSSW system.}

This appendix serves the purpose of summarizing the SCGLE expressions, as well as the asymptotic limits of these and the NE-SCGLE expressions relevant to the construction of glass transition and non-equilibrium glass transition diagrams. We then exemplify the usage of the NE-SCGLE asymptotic expressions by obtaining the non-equilibrium glass transition diagrams of the HSSW fluid system for a variety of attraction lengths.

\subsection{\emph{Equilibrium} SCGLE theory and glass transition diagram.}\label{appendix1.3}

The stationary \emph{equilibrium} limit of Eq. (\ref{relsigmadif2pp}) is $S(k;t\to\infty)\\ =S^{(eq)}(k;n,T)\equiv 
1/n\mathcal{E}(k;n,T)$. In this limit, the NE-SCGLE equations (\ref{dzdtquench})-(\ref{lambdadk}) become
\begin{equation}
\begin{split}
  \Delta \zeta^* (\tau)= \frac{D_0}{24 \pi
^{3}n}
 \int d {\bf k}\ k^2 \left[\frac{ S^{(eq)}(k;n,T_f)-1}{S^{(eq)}(k;n,T_f)}\right]^2 \times \\
F(k,\tau)F_S(k,\tau).
\end{split}
\label{dzdtquencheq}
\end{equation}
\begin{gather}\label{fluctquencheq}
\hat  F(k,z) = \frac{S^{(eq)}(k;n,T_f)}{z+\frac{k^2D^0/S^{(eq)}(k;n,T_f)}{1+\lambda (k)\ \Delta \hat  \zeta^*(z)}},
\end{gather}
and
\begin{gather}\label{fluctsquencheq}
\hat  F_S(k,z) = \frac{1}{z+\frac{k^2D^0 }{1+\lambda (k)\
\Delta \hat \zeta^*(z)}},
\end{gather}
with $\lambda (k)$ being again the phenomenological interpolating function \cite{nescgle1} $\lambda (k)=1/[1+( k/k_{c}) ^{2}]$. 

Eqs.  (\ref{dzdtquencheq})-(\ref{fluctsquencheq}) constitutes the essence of the \textbf{\emph{equilibrium}} \emph{ self-consistent generalized Langevin equation theory} (SCGLE theory). This set of equations are the analog of the MCT dynamic equations (e.g., Eqs. (1) and (3) of Ref. \cite{sperl1}). In both cases, if the equilibrium static structure factor $S^{(eq)}(k;n,T)$ is provided as an input, they allows us to determine the \emph{equilibrium} dynamic properties $\Delta \zeta^* (\tau)$, $F(k,\tau)$, and $F_S(k,\tau)$, as well as the properties that derive from them. These include the mean squared displacement $W(\tau)\equiv  <(\Delta \textbf{r}(\tau))^2> /6$ and the equilibrium mobility $b^{eq}(n,T)$, given by  
\begin{gather}\label{ltsdc}
b^{eq}(n,T)=\left[1+\int_0^{\infty}   \Delta \zeta^* (\tau) d \tau \right]. 
\end{gather}

As first explained in Ref.  \cite{pedroatractivos}, within the equilibrium SCGLE theory the mobility $b^{eq}(n,T)$ and the square localization length $\gamma^{eq}(n,T)\equiv W^{eq}(\tau \to \infty; n,T)$,  play the role of dynamic order parameters. In the equilibrium fluid phase the particles diffuse, and hence, $b^{eq}(n,T)$ is finite and positive while $\gamma^{eq}(n,T)$ is infinite. In contrast, if the system is kinetically arrested, $b^{eq}(n,T)$ vanishes and $\gamma^{eq}(n,T)$ is finite (and is then referred to as the Debye-Waller factor \cite{simmons}). As demonstrated in Ref.  \cite{pedroatractivos}, this dynamic state function can be determined by the solution of the following equation
\begin{equation}
\begin{split}
\frac{1}{\gamma^{eq}(n,T)} =
\frac{1}{6\pi^{2}n}\int_{0}^{\infty }
dkk^4\left[S^{eq} (k;n,T)-1\right] ^{2}\lambda^2(k)\times \\
\frac{1}{\left[\lambda (k)S^{eq} (k;n,T) +
k^2\gamma^{eq}(n,T)\right]\left[\lambda (k) + k^2\gamma^{eq}(n,T)\right]}.
\end{split}
\label{gammaeq}
\end{equation}

Thus, if the  equilibrium static structure factor $S^{eq} (k;n,T)$ is provided, we can scan the state space, either solving the full system of SCGLE equations  (\ref{dzdtquencheq})-(\ref{fluctsquencheq}) to calculate the equilibrium mobility function $b^{eq}(\phi,T)$, or solving this only equation for $\gamma^{eq}(n,T)$. In both cases we can identify the two kinetically-complementary regions (equilibrium ergodic states vs. kinetically arrested states). The application of this protocol to the HSSW model system led to the determination of the solid segment of the dynamic arrest line $T=T_c(\phi)$ of Fig. \ref{fig1}(b), whereas the value of $b^{eq}(\phi,T)$ is represented by a color scale.

\subsection{NE-SCGLE theory and \emph{non-equilibrium} glass transition diagram.}\label{appendix1.4}

Since the equilibrium SCGLE theory is an asymptotic stationary limit, one wonders if the more general NE-SCGLE theory will yield the same dynamic arrest diagram. This question was originally posed and answered in Ref. \cite{nescgle5} Here we report the result in Fig. \ref{fig1}(c), and only mention that the corresponding protocol involves writing the solution of Eq. (\ref{relsigmadif2pp}) for an instantaneous quench to a final $(n,T)$ state point, as $S(k;t)= S^*(k;u(t))$, with $u(t) \equiv \int_0^t b(t')dt'$, and with the function $S^*(k;u)$ given by 
\begin{equation}
\begin{split}
S^*(k;u) \equiv
[n\mathcal{E}(k;n,T)]^{-1}+\left \{ S_i(k)-
[n\mathcal{E}(k;n,T)]^{-1}\right \}\times \\
e^{-2k^2D_0
n\mathcal{E}(k;n,T)u}, 
\end{split}
\label{solsigmadkt}
\end{equation}
with $S_i(k)\equiv S(k;t=0)=S^*(k;u=0)$ being the (arbitrary) initial condition. For each value of $u$ one determines the solution $\gamma^*(u)$ of the equation 
\begin{equation}
\begin{split}
\frac{1}{\gamma^*(u)} =
\frac{1}{6\pi^{2}n}\int_{0}^{\infty }
dkk^4\left[S^*(k;u)-1\right] ^{2}\lambda^2
(k;u)\times \\
\frac{1}{\left[\lambda (k;u)S^*(k;u) +
k^2\gamma^*(u)\right]\left[\lambda (k;u) + k^2\gamma^*(u)\right]}.
\end{split}
\label{nep5ppdu}
\end{equation}
Then, if we find that
$\gamma^*(u)=\infty$ for $0\le u \le \infty$, we conclude that the system will be able to equilibrate after this quench, and hence, that the point $(n,T_f)$ lies in the ergodic region. If, instead, a finite value $u_a$ of the parameter $u$ exists, such that $\gamma^*(u)$ remains infinite \emph{only} within a
finite interval $0\le u < u_a$, the system will no longer
equilibrate, but will become kinetically arrested. 
Thus, if one determines the functions $u_a=u_a(n,T)$ and $\gamma_a^*=\gamma_a^*(n,T)$, one can in principle draw the dynamic arrest diagram, and in Fig. \ref{fig1}(c) we present the result for our HSSW model. The dynamic arrest diagram obtained in this manner will be refer to as \emph{non-equilibrium} glass transition diagram.

\begin{figure}
\begin{center}
\includegraphics[width=.35\textwidth]{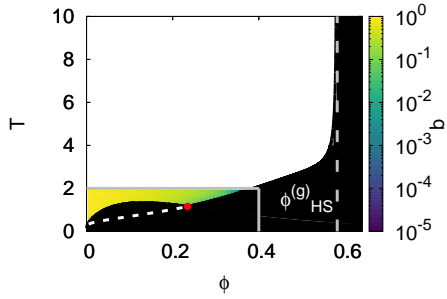}
\caption{Zoom out of the non-equilibrium glass transition diagram of Fig.  \ref{fig1}(c). The region of \ref{fig1}(c) is highlighted by the box at the bottom left corner of the plot. The gray long dashed line represents the Hard Sphere glass transition limit.}
\label{fig6}
\end{center}
\end{figure}

\subsection{Shorter-ranged attractions.}\label{appendix1.5}

Most of the discussion in this paper has referred so far to the portion of state space $(\phi,T)$ around the gas-liquid coexistence region, where we have focused on the interference between gas-liquid phase separation and dynamic arrest. We have thus left out of our discussion other relevant regions, such as the fluid-glass transition line near its hard-sphere limit. To have an idea of the portion of phase space that we have discussed so far, in Fig. \ref{fig6} we reproduce the non-equilibrium diagram of Fig.  \ref{fig1}(c), but in a much wider window, to allow us to indicate the hard-sphere limits for infinite $T$, as a reference to locate the region explicitly studied in this manuscript.

\begin{figure}
\begin{center}
\includegraphics[width=.35\textwidth]{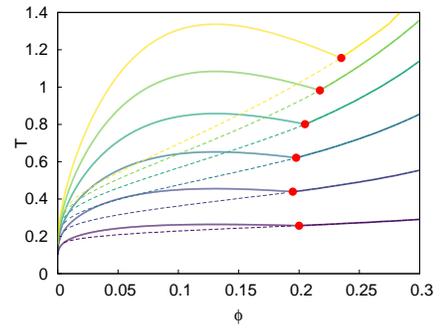}
\caption{Asymptotic non-equilibrium diagram, of the type in Fig.  \ref{fig1}(c), obtained through the NE-SCGLE theory for the HSSW model with well-widths $(\lambda-1) = 0.5$, 0.4, 0.3, 0.2, 0.1, and 0.02. The solid lines represent the boundary between the region of equilibrium fluid states and the region of arrested states. The dashed lines represent the predicted gel-glass transition line. The corresponding bifurcation point is represented by a solid circle.}
\label{fig7}
\end{center}
\end{figure}

Similarly, we have also focused only on the HSSW model with $(\lambda-1) = 0.5$, whose most concise summary is provided by the  illustrative dynamic arrest diagram of Fig. \ref{fig1}(c). The attractive potential range $(\lambda-1) = 0.5$ is realistic for atomic liquids, but not for most of the best studied gel-forming colloidal systems, which frequently involve much shorter-ranged attractions. It is then natural to ask whether one should expect any major differences in the corresponding NE-SCGLE predictions when considering shorter widths of the attractive interactions. Although a thorough analysis of this issue deserves a separate discussion,  for the time being Fig. \ref{fig7}  provides a brief summary of how the asymptotic non-equilibrium phase diagram changes, as  the width of the attractive potential approaches the regime  $(\lambda-1)\ll 0.5$. 

The upper (solid and dashed) lines of Fig. \ref{fig7} reproduces, as a reference,  the  dynamic arrest curves of the asymptotic non-equilibrium phase diagram for $\lambda=1.5$ in Fig.  \ref{fig1}(c). The other solid and dashed lines are the analogous results for a sequence of values of $\lambda$ corresponding to smaller well widths ($(\lambda-1) = 0.4$, 0.3, 0.2, 0.1, and 0.02). In all the cases, the solid lines are composed of two segments. The higher-temperature segment $T=T_c(\phi)$ is the fluid-glass transition curve, which descends from its infinite temperature limit at $\phi_c=0.582$, moving to lower densities as $T$ decreases, until it meets the spinodal line (solid circles), where it bifurcates into the two dynamic arrest lines represented by the low-density segment of the solid line, which coincides with the spinodal line, and the dashed line, which continues the fluid-glass transition line to the interior of the spinodal region as a gel-glass transition. We refer the reader to Ref. \cite{nescgle5} for a thorough explanation of the nature and properties of these transition lines. At this point we only notice the dramatic shrinking, as the width is reduced, of the region bounded from above by the solid line and from below by the dashed line. As discussed in Ref. \cite{nescgle5}, this  is the region where the formation of gels competes with the kinetic pathway to full heterogeneous  gas-liquid equilibrium phase separation (not included in the current version of the NE-SCGLE theory).

\section{Numerical Methods and code availability.} \label{appendix2}

For the data reproducibility we provide access to the ``HS\_HSSW\_quench\_NESCGLE" repository, located at: \url{https://github.com/LANIMFE/HS\_HSSW\_quench\_NESCGLE}. In this repository we solve Eq. (1) just as explained in reference \cite{nescgle3}. For the coupled integro-differential equations (5)-(7) given by the theory, the decimation method described in reference \cite{decimations} is implemented with a maximum of 512 points in an equidistant correlation time, initially starting from $\tau_0=10^{-7}$ with $\Delta\tau_0=10^{-7}$. Additionally, Clenshaw-Curtis quadrature \cite{comp_methods} is being employed for the integral over wave-vectors in (5) with 512 points for $k \sigma \in [ 0:4 ] $ and 512 points for $k \sigma \in [4 :40.96 ] $. Finally, for the integral over correlation time in (4), the Simpson 3/8 \cite{comp_methods} standard integration method is used over equidistant correlation time grids. All other equations not mentioned in this section are directly implemented in a straightforward manner.

The properties of the system used to reproduce the main results of this work are: the square well length $\lambda/\sigma=1.5$, and a grid of final quenching states with final temperatures $T_f\in [ 0.3 : 1.8 ]$, with a resolution of $\Delta T=0.1$, for the isochores $\phi \in [0.01:0.40]$, with a resolution of $\Delta \phi=0.01$. No fixed resolution nor grid is used for $b(t)$ as it can widely vary with system conditions, yet, the program computes the variation of $b(t)$ until $b(t)$ converges up to the seventh significant figure of the final expected mobility in equilibrium conditions or up to $b(t)=10^{-6}$ for dynamical arrest conditions.

\end{document}